\documentclass[aps,prl,twocolumn,groupedaddress]{revtex4-1}
\usepackage{amsmath}
\usepackage{graphicx}
\usepackage{color}
\usepackage{bm}

\newcommand{\bk}{{\bf k}}

\renewcommand{\k}{{\bm{k}}}
\newcommand{\q}{{\bm{q}}}
\renewcommand{\v}{{\bm{v}}}
\renewcommand{\r}{{\bm{r}}}
\newcommand{\R}{{\bm{R}}}
\newcommand{\G}{{\bm{G}}}
\newcommand{\E}{{\mathcal{E}}}

\newcommand{\de}{{\bm{\eta}}}

\newcommand{\cm}{{cm$^{-2}$}}

\def\({\left(}
\def\){\right)}
\def\[{\left[}
\def\]{\right]}

\newcommand{\lr}[1]{ \left( #1 \right) }

\newcommand{\beq} {\begin{eqnarray}}
\newcommand{\eeq} {\end{eqnarray}}

\newcommand{\comment}[1]{}

\begin{document}

\title{Interacting Electrons in Graphene: Fermi Velocity Renormalization
and Optical Response}

\author{T. Stauber$^{1}$}
\author{P. Parida$^{2}$}
\author{M. Trushin$^{3}$}
\author{M. V. Ulybyshev$^{2}$}
\author{D. L. Boyda$^{4,5}$}
\author{J. Schliemann$^{2}$}

\affiliation{$^{1}$ Departamento de Teor\'{\i}a y Simulaci\'on de Materiales, Instituto de Ciencia de Materiales de Madrid, CSIC, E-28049 Madrid, Spain}
\affiliation{$^{2}$Institute for Theoretical Physics, University of Regensburg, D-93040 Regensburg, Germany}
\affiliation{$^3$Department of Physics, University of Konstanz, D-78457 Konstanz, Germany}
\affiliation{$^{5}$Far Eastern Federal University, Sukhanova 8, Vladivostok 690950, Russia}
\affiliation{$^{6}$ITEP, B. Cheremushkinskaya 25, Moscow, 117218 Russia}

\date{\today}
\begin{abstract}
{We have developed a Hartree-Fock theory for electrons on a honeycomb lattice aiming to solve a long-standing problem of the Fermi velocity renormalization in graphene.
Our model employs no fitting parameters (like an unknown band cut-off) but relies on a topological invariant (crystal structure function) 
that makes the Hartree-Fock sublattice spinor independent of the electron-electron interaction.
Agreement with the experimental data is obtained assuming static self-screening including local field effects. As an application of the model, we derive an explicit expression for the optical conductivity and discuss the renormalization of the Drude weight. The optical conductivity is also obtained via precise quantum Monte Carlo calculations which compares well to our mean-field approach.}
\end{abstract}

\maketitle

{\it Introduction.}
The role of Coulomb interactions in graphene is still an open and important 
question,\cite{Kotov12} also in view of the  regime of 
hydrodynamic electron liquids in which the electron-electron interaction represents the 
dominant scattering process.\cite{Bandurin16,Crossno2016} 
The influence becomes especially crucial around the neutrality point, and it has been manifested through the measurement of the effective cyclotron mass,\cite{Elias11} by scanning tunneling spectroscopy,\cite{Li09,Chae12} by direct ARPES of the Dirac cones,\cite{Siegel2011} by quantum capacitance measurements,\cite{Yu13}  and also by Landau level spectroscopy,\cite{Faugeras15} that there is a Fermi velocity renormalization when lowering the electronic density close to half-filling. 

A one-loop renormalization group (RG) and analogous Hartree-Fock (HF) analysis based on the continuous Dirac model predicts  the following scaling behavior \cite{Gonzalez94,Elias11}:
\begin{align}
\frac{v_F^*}{v_F}=1+\frac{\alpha}{4}\ln\frac{\Lambda}{k}\;,
\label{FermiRenorm}
\end{align}
where $\alpha=\frac{1}{4\pi\epsilon_0\epsilon}\frac{e^2}{\hbar v_F}\approx2.2/\epsilon$ is the fine-structure constant of graphene with the bare Fermi velocity $v_F$, $\epsilon$ characterizes the static dielectric environment,
{and $\Lambda$ is the momentum cut-off.}

Eq. (\ref{FermiRenorm}) has been extended by several authors,\cite{Stauber05,Mishchenko07,Sheehy07,Herbut08,Vafek08} but a recent multi-loop expansion claims that perturbation theory may be inadequate 
particularly for suspended graphene.\cite{Barnes14} Nonetheless, within a
non-perturbative functional renormalization group analysis, the perturbative series can be summed up to again yield Eq. (\ref{FermiRenorm}) with almost the same prefactor $\alpha/4$ as obtained
from the HF approach.\cite{Bauer15,Sharma16} This suggests that a self-consistent mean-field theory will contain all the necessary ingredients to address interaction effects even close to the neutrality point. 


The experimental data for the velocity renormalization can be fitted to Eq. (\ref{FermiRenorm}) by adjusting the band cut-off $\Lambda$ as well as the effective dielectric screening constant $\epsilon$.\cite{Elias11} 
Nevertheless, some ambiguities inherent to the renormalization group approach can only be resolved by resorting to a realistic tight-binding Hamiltonian rather than working with an effective low-energy theory.\cite{Abedinpour11} 
This especially holds for the optical conductivity which has been the subject of {persistent} discussion regarding the constant $C$ in the expansion $\sigma^*/\sigma_0=1+C\alpha^*+O({\alpha^*}^2)$,
with $\sigma_0= e^2/4\hbar$ the universal conductivity and $\alpha^*/\alpha=v_F/v_F^*$.
\cite{Herbut08,Mishchenko08,Sheehy09,Juricic10,Sodemann12,Rosenstein13,Teber14,Link16,Boyda16} 
After almost 10 years of debate, one way to resolve this controversy could be an alternative, but well-defined numerical approach which still allows for analytical insight. 
We have therefore performed detailed HF calculations on the honeycomb lattice that hopefully will be able to shed some light on this issue from a different angle.
We complement this with state-of-the-art quantum Monte Carlo calculations which now precisely determine the optical conductivity at energies of the order of the hopping parameter. 

{In contrast to earlier HF calculations preformed for a graphene quantum dot,\cite{Ozfidan15} the Dirac model,\cite{Maxim11}
multilayer graphene,\cite{Trushin2012} and graphene on a lattice\cite{Jung11} 
we now take into account self-screening and finite electronic densities which are shown to be crucial to explain the  experimental data 
{\em without} the need for a fitting parameter.}
We further take advantage of the fact that the HF wave function is independent of the interaction strength even for a lattice model and relate  this to a topological invariant which protects 
the chirality of the Dirac Fermions around the {nodal} points. This reduces the numerical cost and further results in HF equations that in some limits are identical to the ones obtained
from RG equations\cite{Giuliani12} and Hubbard-Stratonovich transformation\cite{Astrakhantsev15}. The knowledge of the HF wave function further enables us to derive analytical expressions for the optical
conductivity in the unscreened case with $C=1/4$, close to the value of Ref. \onlinecite{Juricic10}. Including self-screening, we obtain $C\approx0.05$ for suspended samples in agreement with our Monte Carlo calculations.

The Drude weight renormalization is another motivation of this work, because studies of the electromagnetic response of various classes of correlated electron materials are often based on the f-sum rule.\cite{Basov11} 
Integrating the optical conductivity over the spectral range is then related to the Drude weight $D$ which is independent of 
the interaction in a Galilean invariant system. However, this is not the case for Dirac systems anymore, and electron-electron 
interactions modify the Drude weight in a non-trivial way which is larger than 
the Drude weight of the non-interacting system.\cite{Abedinpour11,Levitov13} 
Sum rule analysis in Dirac systems \cite{Wu15,Post15} have thus to be taken with 
care. Renormalization of the Drude weight is also of interest for plasmonics in Dirac systems as the plasmon energy scales as $\sqrt{D}$.\cite{Stauber14} 
Within our approach, we can analytically discuss the Drude weight for electronic densities close to half-filling.

{\it The model.} We will model interacting Dirac fermions in graphene within a nearest-neighbor tight-binding model using HF theory. 
An important insight of this work is that {the crystallographic structure factor} is a topological invariant which does not depend on the ground-state as long as it obeys the three-fold symmetry of the underlying honeycomb lattice, see Supplementary Information (SI).\cite{SI} 
If we do not allow for chiral\cite{Drut09} or time-reversal\cite{Marino15} symmetry breaking, the HF wave function of the interacting system is then given by the non-interacting wave function. 
This is reminiscent to the absence of wave function renormalization in the RG approach.\cite{Gonzalez94}

In order to consistently include electron-electron interactions, the Coulomb 
potential needs to be periodic as discussed by Jung and MacDonald.\cite{Jung11} This introduces screening of the Coulomb interaction at small distances. 
If one further incorporates screening at large distances due to tight-binding electrons, the interaction potential is not translationally invariant anymore and additional non-diagonal 
local field effects need to be considered, see SI.\cite{SI} To calculate the atomic orbital form factor 
$f(\mathbf{q})=\int d\mathbf{r}e^{-i\mathbf{q}\cdot\mathbf{r}}\left|\zeta({\mathbf r})\right|^2$ with $\zeta({\mathbf r})$ being the one-electron atomic wavefunction,
we take the full angular dependence  of the wave function into account in contrary to Refs. \onlinecite{Jung11} or \onlinecite{Link16}.

\begin{figure}
\includegraphics[width=\columnwidth]{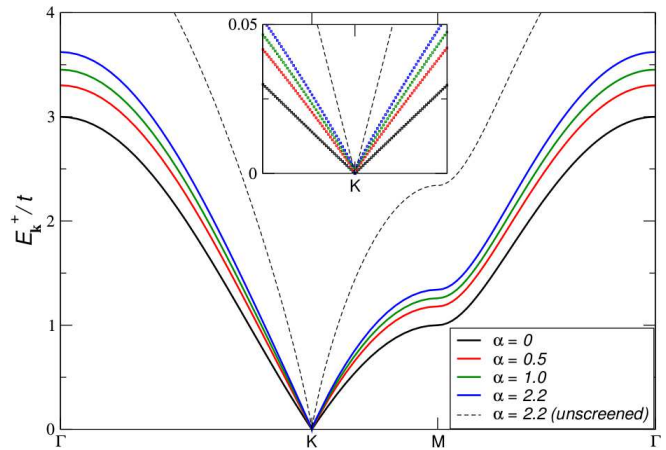}
\caption{(color online) Band structure (undoped) along high symmetry directions for various fine-structure constants $\alpha$ and self-screening (solid lines). For suspended graphene ($\alpha=2.2$, $t=3.1$eV), 
also the unscreened dispersion is shown (dashed line). Inset: Close in of the dispersion around the Dirac cone.
\label{E-k}}
\end{figure}

{\it Hartree-Fock theory.}
As mentioned above, the relative phase between the spinor 
components is a topological invariant and the 
single particle HF Hamiltonian $H_\bk$ for any interaction strength can be written as
\begin{align}
H_\bk=
-\mathcal{E}_\bk\left[\cos(\varphi_\bk)\sigma_x-\sin(\varphi_\bk)\sigma_y\right]\;,
\end{align}
where $e^{i\varphi_\bk}=\phi_\bk/|\phi_\bk|$ with 
the crystallographic structure factor $\phi_\bk =\sum_ie^{i\mathbf {k}\cdot \mathbf {\delta}_i}$, and  the three nearest-neighbor lattice vectors 
$\mathbf{\delta}_i$, $i=1,2,3$. In the SI \cite{SI}, we present the mean-field theory of a more general Hamiltonian which also includes a momentum dependent mass and energy shift for sake of generality.\cite{SI}

The above Hamiltonian is characterized by the renormalized energy dispersion $\mathcal{E}_\bk$ which is determined self-consistently by the following equation:
\begin{align}
\mathcal{E}_\bk=\mathcal{E}_\bk^0+\frac{1}{2A}\sum_{\bk'\in 1.BZ}U(\bk-\bk')e^{i(\varphi_{\bk'}-\varphi_{\bk})}F_{\bk'}\;,
\label{engip}
\end{align}
where we introduced $\mathcal{E}_\bk^0=t\left|\phi_\bk\right|$ as the non-interacting dispersion relation, $t$ is the tunnel-matrix element between nearest carbon atoms and $A$ 
denotes the sample area. Further, we have $F_\bk = n_F(-\mathcal{E}_\bk)- n_F(\mathcal{E}_\bk)$  with the Fermi distribution function 
$n_F(\epsilon)=(e^{\beta(\epsilon-\mu)}+1)^{-1}$ and the chemical potential $\mu$ at a finite temperature $\beta=1/(k_BT)$.
The Coulomb potential conserving the lattice symmetry and including the local field effects reads
\begin{align}
U(\q)&=\sum_{\G,\G'}e^{-i\G'a}f^*(\q+\G) 
f(\q+\G')\notag\\
&\times\left[\delta_{\G,\G'}-v_\G(\q)\chi_{\G,\G'}(\q)\right]^{-1}v_{\G'}(\q)\;,
\label{Ufull}
\end{align}
{where $\G,\G'$ are the reciprocal lattice vectors, $f(\q)$ is the form factor,
$v_\G(\q)=\frac{e^2}{2\epsilon_0\epsilon}\frac{1}{|\q+\G|}$ is the Fourier-transformed screened Coulomb potential, 
and $\chi_{\G,\G'}(\q)$ is the static polarizability matrix with local field effects, see SI \cite{SI}.}
Neglecting the self-consistency by replacing $\mathcal{E}_\bk\rightarrow\mathcal{E}_\bk^0$ on the right hand side of Eq. (\ref{engip}), 
we arrive at the same equation that was obtained from a Hubbard-Stratonovich transformation on the lattice.\cite{Astrakhantsev15} 

In Fig. \ref{E-k}, the renormalized band structure of neutral graphene is shown for $t$=3.1eV ($v_F=10^6$m/s) between the high symmetry points of the Brillouin zone for various coupling constant $\alpha$, 
i.e., for different dielectric environments $\epsilon$. By this, we can discuss suspended graphene ($\epsilon=1$, $\alpha=2.2$), graphene on top of silicon ($\epsilon=2.45$, $\alpha=0.9$) or hBN-encapsulated graphene ($\epsilon=4.9$, $\alpha=0.45$).
The solid lines refer to self-screened interaction which are compared to the dispersion due to bare interaction for $\alpha=2.2$ (dashed line). The inset shows the region close to the Dirac point where only slight deviations from the linear behavior can be seen. 

{\it Fermi velocity renormalization.} At half-filling $\mu=0$, $T=0$ and no self-screening, Eq. (\ref{engip}) is an explicit equation. This yields the analytical expression of Eq. (\ref{FermiRenorm}) when assuming $U(\q)=v_{\G=0}(\q)$ and converting the summation over the Brillouin zone by the Dirac cone approximation. Solving Eq. (\ref{engip}) numerically, we obtain a fit for the cut-off parameter with $\Lambda\approx1.75 \mathrm{\mathring{A}}^{-1}$. This is in contrast to Ref. \onlinecite{Jung11}, 
where  $\Lambda\approx20 \mathrm{\mathring{A}}^{-1}$ was obtained, but agrees well with the usual argument 
of fixing $\Lambda$ by conserving the total number of states in the Brillouin zone when compared to the tight-binding model, yielding $\Lambda\approx1.58 \mathrm{\mathring{A}}^{-1}$.
The precise value of $\Lambda$ depends only weakly on the non-universal short-ranged Coulomb interaction, see SI.\cite{SI}

\begin{figure}
\includegraphics[width=\columnwidth]{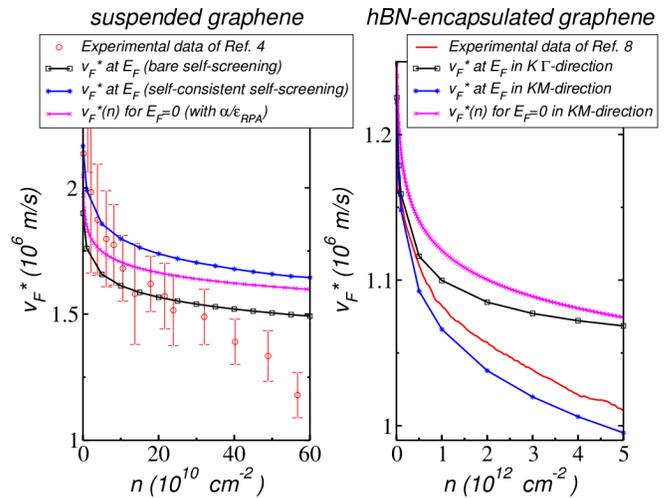}
\caption{(color online) The renormalized Fermi velocity $v_F^*$ for suspended ($\epsilon=1$ and $t=3.1$eV) and hBN-encapsulated ($\epsilon=4.9$ and $t=2.6$eV) graphene. Left hand side for suspended graphene:
The experimental data of Ref. \onlinecite{Elias11} compared to $v_F^*$ at the Fermi surface based on the bare and self-consistent self-screened Coulomb interaction. Right hand side for hBN-encapsulated graphene:
The experimental data of Ref. \onlinecite{Yu13} compared to $v_F^*$ at the Fermi surface in $K\Gamma$ (black squares) and $KM$ (blue stars) direction based on the bare self-screened Coulomb interaction.
In both cases, the result for $v_F^*$ at the neutrality point as function of the electronic density $n$ is also shown based on the unscreened interaction with $\alpha=\alpha/\epsilon^{RPA}$.
\label{vFCompare}}
\end{figure}

For unscreened interaction, the correction to the Fermi velocity $v_F^*/v_F-1$ is proportional to $\alpha$ and the result is scale-independent of the hopping parameter $t$.
At zero doping, self-screening can be incorporated by $\alpha\rightarrow\alpha/\epsilon^{RPA}$ because $\epsilon^{RPA}=1+\frac{\pi}{2}\alpha$ is momentum independent
within the Dirac-cone approximation \cite{Kan04}. This yields good agreement with the experimental data of Ref. \onlinecite{Elias11} for small densities $n\lesssim 20\times10^{10}$\cm {\em without} 
the need for any fitting parameter, see magenta curve of Fig. \ref{vFCompare}.

For densities $n\gtrsim20-60\times10^{10}$\cm, there is a decrease of the Fermi velocity which cannot be accounted for by the results for neutral graphene.
Since it might be due to screening at finite densities, we incorporated the momentum dependent polarization function as outlined above. Even though the renormalized Fermi velocity 
now depends on $\alpha$ as well as on $t$ in a non-trivial way, in the asymptotic limit it becomes independent of $t$ for $\mu=0$. 

On the left hand side of Fig. \ref{vFCompare}, we show the solution of Eq. (\ref{engip}) using the bare (black squares) and the self-consistent (blue stars) polarization function. 
The bare solution agrees well with the experimental data {for suspended graphene} up to {$n \lesssim 40\times10^{10}$\cm}, 
but {at higher densities} the experimental data drops 
whereas the theoretical value remains approximately constant (on a logarithmic scale).
This has to be contrasted with the experimental data for hBN-encapsulated graphene,\cite{Yu13} where good agreement is obtained over the whole density range up to $n\sim5\times10^{12}$\cm, see Fig. \ref{vFCompare} (right). 
For the particular choice of the bare hopping amplitude in the two cases, i.e., $t=3.1$eV and $t=2.6$eV, 
respectively, we were guided by the original Refs. \cite{Elias11} and \cite{Yu13} where similar values were used, see also Ref. \cite{Rei02}. 

{\it Optical response.} Let us now turn to the interaction effects on the optical response, first discussed for Dirac electrons in Ref. \onlinecite{Stauber08} in the case of electron-phonon coupling. To do so, we will couple the gauge field via the Peierls substitution by replacing 
$\bk \rightarrow \bk +\frac{e}{\hbar}\mathbf{A}$ in the mean-field Hamiltonian 
$H_\bk$. This procedure provides the correct vertex correction such that the 
optical f-sum rule is satisfied.  

Since the HF wave function is known, the optical conductivity can be deduced from the non-interacting tight-binding model\cite{Peres08} by replacing the bare dispersion by the renormalized one. 
For momenta close to the Dirac point, the dispersion is {isotropic}  with $v^*_F/v_F=1+C(\alpha)\alpha\ln\Lambda/k$. 
We then obtain for small frequencies $\omega \ll t/\hbar$ the following result:\cite{SI}
\begin{flalign}
\frac{\sigma^*}{\sigma_0}=1+C(\alpha)\alpha\frac{v_F}{v_F^*}\;,   
\label{SigmaDirac}
\end{flalign}
where $\sigma_0=\frac{e^2}{4\hbar}$ denotes the universal conductivity. For unscreened interaction, we obtain
{\em {explicitely}} $C=1/4$ {(see SI\cite{SI})} as mentioned in the introduction. 
This compares well with $C\approx0.26$ obtained in Refs. \onlinecite{Juricic10,Rosenstein13}.

\begin{figure}
\includegraphics[width=\columnwidth]{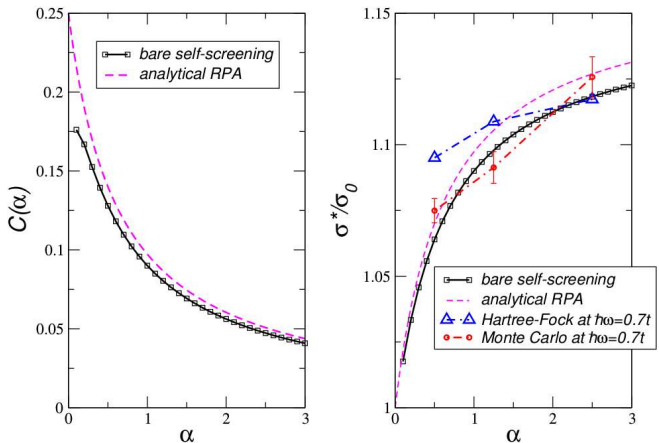}
\caption{(color online)  Left: The correction to the optical conductivity $C(\alpha)$ of Eq. (\ref{SigmaDirac}) compared to the result expected from the RPA, i.e., $[4\epsilon^{RPA}]^{-1}=[4(1+\frac{\pi}{2}\alpha)]^{-1}$. Right:
The same for the conductivity of Eq. (\ref{SigmaDirac}) with $v_F^*/v_F=1$ compared to the conductivity at $\hbar\omega=0.7t$ obtained from Hartree-Fock (blue triangle) and quantum Monte Carlo (red circle) calculations.
\label{C}}
\end{figure}

For the self-screened interaction, $C\rightarrow C(\alpha)$ becomes a function with $C(\alpha\to0)\to0.25$ (unscreened limit) and $C(\alpha=2.5)\approx0.05$ for suspended graphene with $t=2.7$eV. 
It is interesting to note that the universal factor of the scaling law $C(\alpha)$ is independent of all considered hopping matrix elements from $t=2.6$eV to $t=3.1$eV, and it compares well if self-screening is incorporated via RPA within the Dirac cone approximation,
i.e., $C(\alpha)=[4(1+\frac{\pi}{2}\alpha)]^{-1}$. This is shown on the left of Fig. \ref{C}; assuming a small, but finite electronic density would further decrease the constant $C(\alpha)$.

On the right hand side, we plot the same for the conductivity of Eq. (\ref{SigmaDirac}) with $v_F^*/v_F=1$. This is compared to the conductivity at $\hbar\omega=0.7t$ as obtained from HF and quantum Monte Carlo calculations, see SI \cite{SI}. We obtain good agreement between the two approaches for suspended graphene which justifies our mean-field approach.

{\it Drude weight.} In a Galilean invariant system, the Drude weight $D$ is independent of 
the interaction. However, this is not the case for Dirac systems and electron-electron 
interactions modify the Drude weight in a non-trivial way which is larger than 
the Drude weight of the non-interacting system. \cite{Abedinpour11,Levitov13}

Making use of the fact that the HF wave function is given by the non-interacting wave function, one can obtain an analytical expression for the Drude weight from the optical $f$-sum rule {(see SI\cite{SI})}: 
\begin{flalign}
D_{ii} = \left(\frac{e }{\hbar}\right)^2\frac{g_s}{A}\sum_{\bk\in 1.BZ,s=\pm} 
s[\partial_{k_i}^2\mathcal{E}_{\bk}]n_F(s\mathcal{E}_{\bk})\;,
\label{drude}
\end{flalign}
\noindent
with $g_s=2$ the spin degeneracy, and $i=x,y$. 
{This expression for the Drude weight in the presence of Coulomb interactions 
generalizes the known result for non-interacting systems:
$D=e^2n/m$ for Schr\"odinger particles with the density $n$ and mass $m$ and
$D=g_s(e/h)^2\pi\mu$ for for Dirac particles in graphene.\cite{Stauber14}}
As it might have been expected, within our mean-field theory the interacting electrons behave as independent quasi-particles with renormalized dispersion 
$\lambda\mathcal{E}_{\bk}$ and we obtain\cite{SI} 
\begin{align}
\frac{D^*}{D}=\frac{v_F^*}{v_F}\;.
\end{align}
A similar relation was obtained in Ref. \onlinecite{Abedinpour11} in the case of unscreened interactions.
Changes due to trigonal warping and finite temperature can also be discussed by our approach.

{\it Summary.}
We presented a realistic tight-binding approach to the electron band structure of graphene renormalized by the Coulomb interaction.
We identified a topological invariant which leaves the HF wave function unchanged even in the presence of
the Coulomb interaction and found analytical expressions for the optical conductivity as well as for the Drude weight.
By this, we were able to link our findings  to the measured optical conductivity which shows only little renormalization due to self-screened interactions.
Precise Monte Carlo calculations yield good agreement for suspended samples and support our mean-field approach. We also show that the Fermi velocity and Drude weight renormalization are the same according to the expectations of a mean-field theory. 

Our results compare well with experiments for suspended as well as for hBN-encapsulated graphene without invoking fitting parameters.
But in the case of suspended graphene, we were not able to account for the velocity renormalization in the case of larger densities $n\gtrsim40\times10^{10}$\cm. This effect cannot be explained by our tight-binding model and we expect the influence of ripples and corrugations, partially due to the applied gate, to be responsible for this effective screening of the long-ranged Coulomb interaction. This would also imply the absence of interaction renormalization of (high-density) plasmons in suspended graphene.

{\it Acknowledgements}. We acknowledge interesting discussions with Igor Herbut, Jeil Jung, Peter Kopietz, Vieri Mastropietro, and Anand Sharma. We also thank Kostya Novoselov for providing us with the experimental data. This work has been supported by Spain's MINECO under Grant No. FIS2014-57432-P, by the Comunidad de Madrid under Grant No. S2013/MIT-3007 MAD2D-CM, and by Germany's Deutsche Forschungsgemeinschaft (DFG) via SFB 689. The work of M. Ulybyshev was supported by DFG grant BU 2626/2-1. D. Boyda  acknowledges the support by RFBR Grant No. 16-32-00362-mol-a.

\begin{widetext}
{\bf Supplementary Information}
\section{The non-interacting tight-binding model}    
We will model interacting Dirac fermions in graphene within a nearest-neighbor 
tight-binding model. The honeycomb lattice of graphene is a non-Bravais 
bipartite lattice, which
means that it consists of two interpenetrating sublattices, each of them 
forming a
triangular Bravais lattice. The lattice vectors are chosen as follows:
\begin{equation}
\bm {a}_1 = \frac{ a}{2}\left( 3 ,  \sqrt{3}\right), ~~~~ \; \bm {a}_2 
= \frac{ a}{2}\left( 3 , - \sqrt{3}\right),
\label{graphlattice}
\end{equation} 

\noindent where a $\sim 1.42\, \mathrm{\mathring{A}}$ is the C-C covalent bond length.
The three nearest-neighbor vectors in real space are given by
\begin{equation}
\bm {\delta}_1 = \frac{a}{2}\left(-1, \sqrt{3}\right)\;,\; 
\bm{\delta}_2 = \frac{a}{2}\left(-1, -\sqrt{3}\right)\;,\; 
\bm{\delta}_3 = a\left(1,0\right)\;.
\label{nnbr}
\end{equation}
\noindent The reciprocal lattice vectors ${\bm b}_1$ and ${\bm b}_2$ defined by 
the condition 
${\bm a}_i\cdot{\bm b}_j = 2 \pi \delta_{ij}$
are then given by
\begin{align}
\bm {b}_1 = \frac{2\pi}{3a}\left(1,\sqrt{3}\right)\;,\; 
\bm {b}_2 = \frac{2\pi}{3a}\left(1,-\sqrt{3}\right). 
\label{recvec}
\end{align}
Within the nearest-neighbor tight-binding model, the general Bloch basis state
is given by
\begin{align}
\bm {\Psi}_{{\bk}\lambda} (\bm {r}) = \frac{1}{\sqrt{N_c}}
\sum_{j}e^{i\bm{k}\cdot\left(\bm {R}_j + \bm {\eta}_\nu\right)}
\zeta(\bm {r}-\bm{R}_j-\bm {\eta}_\nu)\xi_\sigma,    
\label{bloque}
\end{align}
where $N_c$ is the number of unit cells, $\xi_\sigma$ denotes the spin part of 
the wave function, $\zeta$ the one-electron atomic wavefunction 
($p_z$ orbital of carbon) at 
$\bm {R}_j+ \bm{\eta}_\nu$, 
$\bm \eta_\nu$ is the position of sublattice $\nu$ in the crystallographic 
basis, and $\lambda=(\nu,\sigma)$ shall include sublattice and spin degrees of 
freedom. In the following, we will mostly choose $\bm{\eta}_{a}= (0,0)$, $\bm{\eta}_{b} = (a,0)$.
  
The free tight-binding Hamiltonian can then be written as
\begin{equation}
H_\bk^0 =
-t |\phi_\bk|\left( \begin{array}{cc} m_\bk^0 & e^{i\varphi_\bk} \\
e^{-i\varphi_\bk}& -m_\bk^0 \end{array} \right)+E_\bk^01_{2\times2},
\label{tightbinding}
\end{equation}
where $t$ = 3.1eV, $\phi_\bk =\sum_{i=1,2,3}e^{i\bm {k}\cdot \bm {\delta}_i}$ 
the structure factor, and $e^{i\varphi_\bk}=\phi_\bk/|\phi_\bk|$. We also 
included a mass term and a constant energy term for sake of generality. The 
eigenenergies read
\begin{flalign}
\mathcal{E}_{\bk}^{0,\pm}  =  E_\bk^0 \pm t\left|\phi_\bk\right|\sqrt{1+ 
{m_\bk^0}^2}. 
\label{eigenenergies0}
\end{flalign}
The eigenvectors are given by
\begin{flalign}
|\psi_{\bk}^-\rangle =
 \left(\begin{array}{c}\cos\frac{\vartheta_{\bk}^0}{2}  \\
 \sin\frac{\vartheta_{\bk}^0}{2 }e^{-i\varphi_\bk}\end{array} \right)\;,\;
|\psi_{\bk}^+\rangle=
 \left(\begin{array}{c}\sin\frac{\vartheta_{\bk}^0}{2}  \\
-\cos\frac{\vartheta_{\bk}^0}{2}e^{-i\varphi_\bk}\end{array} \right) \;, 
\label{EigenVectors}
\end{flalign}
with $\cos\vartheta_\bk^0=m_\bk^0/\sqrt{1+ {m_\bk^0}^2}$ and 
$\sin\vartheta_\bk^0=1/\sqrt{1+ {m_\bk^0}^2}$. We note that by defining the Bloch state as usual, i.e., $\bm {\Psi}_{{\bk}\lambda} (\bm {r}+\bm{R}_i)=e^{i\bk\cdot\bm{R}_i}\bm {\Psi}_{{\bk}\lambda} (\bm {r})$, we arrive at the eigenvectors of the {Hamiltonian} with a relative phase $e^{-i\varphi_\bk}$ between the first and second spinor component.

\section{Mean-Field theory}
Let us introduce the Hamiltonian for the electron-electron interaction as
\begin{align}
V=\frac{1}{2}\sum_{i,j;\lambda,\lambda'}c_{i\lambda}^\dagger c_{j\lambda'}^\dagger\langle i\lambda,j\lambda'|V|i\lambda,j\lambda'\rangle c_{j\lambda'}c_{i\lambda}\;.
\end{align}
We define the Fourier transformation as follows:
\begin{align}
c_{i\lambda}^\dagger=\frac{1}{N_c}\sum_{\k\in1.BZ}e^{-i\bk\cdot(\R_i+\eta_\nu)}c_{\k\lambda}^\dagger\;,\;
c_{\k\lambda}^\dagger=\frac{1}{N_c}\sum_{i}e^{i\bk\cdot(\R_i+\eta_\nu)}c_{i\lambda}^\dagger\;.
\end{align}
This is consistent with Eq. (\ref{bloque}), i.e., ${\Psi}_{{\bk}\lambda} (\bm {r})=\langle\r|c_{\k\lambda}^\dagger|0\rangle$. With the Coulomb propagator $U_\q^{\lambda,\lambda'}$ defined in the next section, we then have
\begin{align}
\label{Interaction}
V=\frac{1}{2A}\sum_{\k,\k',\q}\sum_{\lambda,\lambda'}U_\q^{\lambda,\lambda'}c^\dagger_{\bk+\q\lambda}c^\dagger_{\bk'-\q\lambda'}c_{\bk'\lambda'}c_{\bk\lambda}
\end{align}
where $A=A_cN_c$ is the system area with $A_c=3\sqrt{3}a^2/2$.

The interaction term within the mean-field approximation reads
\begin{align}
H_\bk^{ee} & = \sum_{\lambda,\lambda'}U^{\lambda\lambda'}_0\frac{1}{A}\sum_{\bk'}
\langle c^\dagger_{\bk'\lambda'}c_{\bk'\lambda'}\rangle c^\dagger_{ \bk\lambda}c_{ 
\bk\lambda} \notag \\
& -\frac{1}{A}\sum_{\bk'\lambda,\lambda'}U^{\lambda\lambda'}_{\bk- \bk'}
\langle c^\dagger_{\bk'\lambda'}c_{\bk'\lambda}\rangle c^\dagger_{\bk\lambda}c_{
\bk\lambda'}.  
\label{HF}
\end{align}
The two terms are the 
Hartree and Fock (exchange) term, respectively.  
In the following, we will only consider the exchange interaction of the HF 
Hamiltonian as the Hartree term is neutralized by a positive background.   

We numerically solved the total Hamiltonian $H=\sum_\bk H_\bk$, $H_\bk = H_\bk^0 + H_\bk^{ee}$, and found that the in-plane (azimuthal) angle of the pseudospin phase $\varphi$ is not changed by the interaction strength. This can also be shown analytically, see below. The relative phase between the spinor 
components can thus be seen as a topological invariant and with $\phi_\k^{ee}=\langle c_{\k,b,\sigma}^\dagger c_{\k,a,\sigma}\rangle/2\propto\phi_\k$, the single particle $H_\bk$ for any interaction strength can be written as
\begin{equation}
H_\bk=
-t |\phi_\bk^{ee}|\left( \begin{array}{cc} m_\bk & e^{i\varphi_\bk} \\
e^{-i\varphi_\bk}& -m_\bk \end{array} \right)+E_\bk1_{2\times2}\;.
\label{MeanField}
\end{equation}   
This parametrization will allow for a more efficient numerical analysis, crucial in order to address the physics close to the neutrality point. The eigenenergies of the interacting system now read
\begin{flalign}
\mathcal{E}_{\bk}^{\pm}  =  E_\bk \pm t\left|\phi_\bk^{ee}\right|\sqrt{1+ 
m_\bk^2}
\label{eigenenergies}
\end{flalign}
and the eigenvectors are again given by Eq. (\ref{EigenVectors}) after replacing $\vartheta_\bk^0\rightarrow\vartheta_\bk$ where
$\cos\vartheta_\bk=m_\bk/\sqrt{1+m_\bk^2}$ and 
$\sin\vartheta_\bk=1/\sqrt{1+m_\bk^2}$.

In the above equations, we have introduced the renormalized energy dispersion $|\phi_\bk^{ee}|$ if $m_\bk=0$ and $E_\bk=0$, the total dimensionless, $\bk$-dependent mass $m_\bk$, and the total $\bk$-dependent energy shift $E_\bk$. These quantities are 
determined {self-consistently} by the following relations:
\begin{flalign}
t\left|\phi_\bk^{ee}\right|&=t\left|\phi_\bk\right|+ \frac{1}{2A}\sum_{\bk'}U^{12}_
{\bk-\bk'} \frac{e^{i(\varphi_{\bk'}-\varphi_{\bk})} 
}{\sqrt{1+m_{\bk'}^2}}F_{\bk'}^-\;, \label{condition1} \\ 
t\left|\phi_\bk^{ee}\right|m_\bk&=t\left|\phi_\bk\right|m_\bk^{0}+ \frac{1}{2A}
\sum_{\bk'}U^{11}_{\bk-\bk'} \frac{m_{\bk'}}{\sqrt{1+ m_{\bk'}^2}}F_{\bk'}^-\;, 
  \label{condition2}\\
E_\bk&=E_\bk^0-\frac{1}{2A}\sum_{\bk'}U^{11}_{\bk-\bk'}F_{\bk'}^+ \;.
\label{condition3}
\end{flalign}
Above, we defined $F_\bk^\pm = f(\E_{\bk}^-)\pm f(\E_{\bk}^+)$  with 
$f(\epsilon)=(e^{\beta(\epsilon-\mu)}+1)^{-1}$ the Fermi distribution function 
and $\mu$ the chemical potential at a finite 
temperature $\beta=1/(k_BT)$.  

During the iteration process to obtain self-consistency, the chemical potential relative to the neutrality point will be kept constant. For $\mu=0$, Eq. (\ref{condition3}) becomes trivial because $\sum_{\bk'}U^{11}_{\bk-\bk'}=0$. If we also neglect the self-consistency, we arrive at the same set of equation if one-loop corrections to the electron propagator and to the interaction potential are considered.\cite{Astrakhantsev15}  

\section{Screening and local field effects}
We will now specify the Coulomb propagator $U^{\nu,\nu'}_\q$ defined in Eq. (\ref{Interaction}). If one incorporates screening due to tight-binding electrons, the interaction potential is not translationally invariant anymore and we have instead 
($\lambda=(\nu,\sigma)$)
\begin{align}
\langle i\lambda,j\lambda'|V|i\lambda,j\lambda'\rangle=\int d^2r_1\int 
d^2r_2|\zeta(\bm{r}_1-\R_i-\de_\nu)|^2V(\bm{r}_1,\bm{r}_2)|\zeta(\bm{r}_2-\R_j-\de_{
\nu'})|^2\;.
\end{align}
With the condition $V(\bm{r}_1,\bm{r}_2)=V(\bm{r}_1+\R_i,\bm{r}_2+\R_i)$, we obtain the following representation:\cite{giu05}
\begin{align}
V(\bm{r}_1,\bm{r}_2)=\frac{1}{A^2}\sum_{\q\in 
1.BZ,\G_1,\G_2}e^{i(\q+\G_1)\bm{r}_1}e^{-i(\q+\G_2)\bm{r}_2}V(\q+\G_1,\q+\G_2)\;,
\end{align}
where the sum runs over all reciprocal lattice vectors $\G=n\bm {b}_1+m\bm {b}_2$ with integers $n,m\in\bm{Z}$.
The interaction potential $H^{ee}$ of Eq. (\ref{Interaction}) is then defined with the Coulomb propagator given by
\begin{align}
U^{\nu,\nu'}_\q=\frac{1}{A}\sum_{\G,\G'}e^{i\G\de_\nu}e^{-i\G'\de_{\nu'}}f^*(\q+\G) 
f(\q+\G')V(\q+\G,\q+\G')
\label{UfullSI}
\end{align}
with $f(\q)=\int d^2r|\zeta(\bm{r})|^2e^{-i\q\cdot\bm{r}}$ and where we used 
$c_{\k+\G,\lambda}=e^{i\G\cdot\de_\nu}c_{\k,\lambda}$.
We note that the definition is independent of the choice of the crystallographic basis $\de_\nu$ and that $U^{\nu,\nu'}_\q=U^{\nu',\nu}_{-\q}$. It can also be shown that $U^{1,1}_\q=U^{2,2}_{\q}$ expected from the equivalence of the two sub-lattices.

The screened potential shall be calculated within the RPA-approximation:
\begin{align}
V(\q+\G,\q+\G')=A\left[\delta_{\G,\G'}-v_\G(\q)\chi_{\G,\G'}(\q)\right]^{-1}v_{
\G'}(\q)\;,
\end{align}
with $v_\G(\q)=v(\q+\G)=\frac{e^2}{2\epsilon_0\epsilon |\q+\G|}$.
The (dynamical) response function is given by 
\begin{align}
\chi_{\G,\G'}(\q,\omega)&=\frac{1}{A}\sum_{\k,s,s'}\frac{n_F(\mathcal{E}_{\k}^s)-n_F(\mathcal{E}_{\k+\q}^{s'})}{\mathcal{E}_{\k}^{s}-\mathcal{E}_{\k+\q}^{s'}
+\hbar\omega+i0}
\langle\k,s|e^{-i(\q+\G)\bm{\hat r}}|\k+\q,s'\rangle\langle\k+\q,s'|e^{i(\q+\G')\bm{\hat r}}|\k,
s\rangle\;.
\end{align}
The eigenfunctions are given by 
\begin{align}
\langle\bm{r}|\k,s\rangle=\sum_{\nu=a,b}\psi_{\nu,\k,s}\Psi_{\k,\lambda}(\bm{r})
\end{align}
with the general Bloch eigenstate $\Psi_{\k,\lambda}(\bm{r})$ given in Eq. (\ref{bloque}) and $\psi_{a,\k,-}=\cos\frac{\vartheta_{\bk}}{2}$, 
$\psi_{b,\k,-}=\sin\frac{\vartheta_{\bk}}{2 }e^{-i\varphi_\bk}$, 
$\psi_{a,\k,+}=\sin\frac{\vartheta_{\bk}}{2 }$, and 
$\psi_{b,\k,+}=-\cos\frac{\vartheta_{\bk}}{2}e^{-i\varphi_\bk}$. The matrix 
overlap function is given by 
\begin{align}
\langle\k,s|e^{-i(\q+\G)\bm{\hat r}}|\k+\q,s'\rangle=f(\q+\G)\sum_{\nu=a,b}\psi_{\nu,\k,s}^*\psi_{\nu,\k+\q,s'}e^{-i\G\cdot\de_\nu}\;.
\end{align}
Finally, we note that due to time-reversal {symmetry} we have $\chi_{\G,\G'}(\q,\omega)=\chi_{\G',\G}(-\q,\omega)$.


We note that the condition
\begin{align}
U^{\nu,\nu'}_{\q+\G}=e^{-i\G\cdot(\de_\nu-\de_{\nu'})}U^{\nu,\nu'}_{\q}
\end{align}
is fulfilled in all cases which is needed to fulfill the self-consistency equations as we will show in the next section. 

\subsection{Dirac cone approximation}
For small in-plane momentum $q$, we can use the Dirac cone approximation for the evaluation of the sum over the Brillouin zone. This can be done analytically and we get for the doped case:
\begin{align}
\label{chiLocalField}
\chi_{\G,\G'}(q)&=-\frac{g_sg_vk_F}{2\pi v_F\hbar}f(\q+\G)f(\q+\G')\\\notag
&\times\Bigg[\gamma_1+\left(\frac{q}{4k_F}\gamma_2\arccos(\frac{q}{2k_F})-\gamma_3\frac{1}{2}\sqrt{1- (\frac{2k_F}{q})^2}\right)\Theta(\frac{q}{2k_F}-1)\\\notag
&+\frac{q}{4k_F}\gamma_2\left(\arccos(\frac{q}{2\Lambda})-\frac{\pi}{2}\right)+\gamma_4\sqrt{(2\Lambda)^2-q^2}/k_F\Bigg]
\end{align}
with $\gamma_1=(e^{-i\G\cdot\eta}+e^{i\G'\cdot\eta})/2$, $\gamma_2=(e^{-i\G\cdot\eta}+e^{i\G'\cdot\eta})/2-(1-e^{-i\G\cdot\eta})(1-e^{i\G'\cdot\eta})/4$, $\gamma_3=(1+e^{-i\G\cdot\eta})(1+e^{i\G'\cdot\eta})/4$, $\gamma_4=(1-e^{-i\G\cdot\eta})(1-e^{i\G'\cdot\eta})/8$.
For the undoped case, this reduces to
\begin{align}
\chi_{\G,\G'}(q)&=-\frac{g_sg_v}{2\pi v_F\hbar}f(\q+\G)f(\q+\G')
\Bigg[\gamma_2\frac{q}{4}\arccos(\frac{q}{2\Lambda})+\gamma_4\sqrt{(2\Lambda)^2-q^2}\Bigg]\;.
\end{align}

The band-cut off $\Lambda$ shall not be confused with the one used in the scaling law of the Fermi velocity and can be obtained from the exact numerical solution. We get $\Lambda=2.29\, \mathrm{\mathring{A}}^{-1}$.
\section{Topological invariant leads to phase locking between the spinor components}
Here, we show that the phase between the components of the spinor is invariant with respect to the Coulomb interaction. This property is due to the boundary condition $c_{\k+\G,\lambda}=e^{-i\G\cdot\de_\nu}c_{\k,\lambda}$ and the underlying three-fold and at the Dirac point rotational symmetry, i.e., we neglect possible symmetry breaking due to interaction.

Let us define $\phi_\k^{ee}=\langle c_{\k,b,\sigma}^\dagger c_{\k,a,\sigma}\rangle/2$. The boundary condition implies
\begin{align}
\label{BC}
\phi_{\k+\G}^{ee}=e^{-i\G\cdot(\bm{\eta}_{a}-\bm{\eta}_{b})}\phi_\k^{ee}=e^{iG_xa}\phi_\k^{ee}\;.
\end{align} 
where we choose $\bm{\eta}_{a}= (0,0)$, $\bm{\eta}_{b} = (a,0)$.

Due to the three-fold symmetry, we have $\phi_\k^{ee} = \phi_{O\k}^{ee}  = \phi_{O^2\k}^{ee}$  where $O$ is a rotation about  the angel $\varphi=2\pi/3$. Thus we can formally set $f_1(\k)  =\phi_\k^{ee}/3$,  $f_2(\k) = \phi_{O\k}^{ee}/3$, $f_3=\phi_{O^2\k}^{ee}/3$ and write $\phi_\k^{ee}=\sum_{i=1..3}f_i(\k)$. At the Dirac point, the system is further rotationally invariant and we can set $f_i(\k)=f(\v_i\cdot\k)$, with an arbitrary function $f$ and $\v_i$ being the three vectors related by a rotation of $\varphi=2\pi/3$. In order to obtain the boundary condition Eq. (\ref{BC}), the arbitrary function must be related to the exponential and we thus have $f(x)=Ae^{ix}$ and $\v_i={\bm \delta}_i$ the {nearest-neighbor} vectors with an arbitrary constant $A$.

We thus have $\phi_\k^{ee}\propto\sum_{i}e^{i{\bm 
\delta}_i\cdot\k}=\phi_\k={}_0\langle c_{\k,b,\sigma}^\dagger c_{\k,a,\sigma}\rangle_0/2$. The crystallographic structure factor in the case of the non-interacting system can thus be obtained from general symmetry arguments with the same boundary conditions as in Eq. (\ref{BC}) and it is independent of the interaction or other perturbations which preserve the basic symmetries. 

Finally, let us note that Eq. (4.21) of Ref. \cite{Giuliani12} comes to the same conclusion based on the Feynman graphs with two external fermionic lines with the same wave vector $\bk$ and a more formal proof of this relation can be found in App. C of the same reference. 
 
\section{Derivation of the form factor}
Including lattice scale effects, the explicit form of $f(q)$ can be 
calculated 
considering $\zeta(r)$ to be a hydrogenic 2$p_z$ orbital type, 
$\zeta(r,\vartheta)= 
\frac{1}{4\sqrt{2\pi}}(\frac{Z}{a_0})^{3/2}\frac{Zr}{a_0}e^{\frac{-Zr}{2a_0}}
\cos(\vartheta)$ with $Z$ the effective (screened) atomic charge.
By taking the Fourier transformation of the charge distribution, the form factor is defined by 
\begin{flalign}
f(\textbf{q})=\int 
d\textbf{r}e^{-i\textbf{q}.\textbf{r}}\left|\zeta(\textbf{r})\right|^2\;.
\end{flalign}
We choose $\textbf{q}$ to be in the $xy$-plane with $\textbf{q}$ = q(1,0,0) and define $\tilde{a}_0=\frac{a_0}{Z}$. This yields
\begin{align}
f(\textbf{q}) & = \frac{1}{32 \pi \tilde{a}_0^3}\int d\textbf{r}
\left(\frac{r}{\tilde{a}_0}\right)^2 e^{-i\textbf{q}\cdot\bf{r}}e^{-\frac{r}{\tilde{a}_0}}
\cos^2\vartheta \notag \\
& = \frac{1}{32 
\pi}\int_0^{\infty}dxx^4e^{-x}\int_0^{\pi}
d\vartheta\sin\vartheta\cos^2\vartheta \notag\int_0^{2\pi}d\phi\sum_{n=0}^{\infty}\frac{(-iq\tilde{a}
_0\sin\vartheta\cos\phi)^n}{n!} \notag \\
& = \sum_{n=0}^{\infty}(q\tilde{a}_0)^{2n}(-1)^n\frac{(n+1)(n+2)}{2} 
\end{align}
We thus obtain the final result:
\begin{align}
f(q)= \frac{1}{(1+q^2\tilde{a}_0^2)^3}
\end{align}

With $\langle r^2\rangle=30\tilde{a}_0^2$, we can reproduce the covalent bond radius of carbon by choosing $\tilde{a}_0=a/2/\sqrt{30}$. A larger effective radius could also consider screening effects from the sp$^2$-orbitals and we will choose $\tilde{a}_0=3a/2/\sqrt{30}$.\cite{Jung11} The short-distance cut-off has only little influence on the scaling behaviour of the Fermi velocity, i.e., the cut-off parameter only slightly increases with increasing short-distance screening corresponding to the absolute band width with 
$\Lambda=1.75\, \mathrm{\mathring{A}}^{-1}$ and $\Lambda=1.82\, \mathrm{\mathring{A}}^{-1}$ for the two parameters mentioned above. 

\section{Optical response} 
Let us now discuss the interaction effects on the optical response. For this, we will couple the gauge field employing the Peierls substitution by replacing 
$\bk \rightarrow \bk +\frac{e}{\hbar}\bm{A}$ in the mean-field Hamiltonian 
$H_\bk$. This procedure provides the correct vertex correction such that the 
optical f-sum rule is satisfied.  

Assuming a linearly polarized light along the 
$i$-direction with $i=x,y$, the current operator reads
\begin{flalign}
\label{current}
j_i = - \frac{\partial H}{\partial A_i }= j_i^P + j_i^DA_i + \mathcal{O}(A_i^2),
\end{flalign}
where $j_i^P$ and $j_i^D$ are the paramagnetic and diamagnetic contribution, 
respectively.

The real part of the conductivity $\sigma_{ii}(\omega)$ represents the optical 
absorption of the light. Using the Kubo formula within the linear response 
formalism, $\Re{\sigma}_{ii}$ can be split into two terms containing a regular 
part and a delta singularity
\begin{align}
\Re{\sigma}_{ii}(\omega) = \pi D_{ii}\delta(\omega) + \sigma_{ii}^{reg}(\omega),
\end{align}
where $D_{ii}$ is the Drude weight corresponding to the charge stiffness. The 
regular part of the conductivity reads for $\omega>0$
 \begin{flalign}
\sigma_{ii}^{reg}(\omega) &= \left(\frac{e}{\hbar}\right)^2\frac{g_s\pi}{A\hbar\omega} 
\sum_{\bk} \left|P_\bk^i\right|^2F_{\bk}^- \delta(\E_\bk^- - \E_\bk^+ + 
\hbar\omega) \;,  
\label{regular1}
\end{flalign} 
\noindent
where $P_\bk^i = \langle\psi_{\bk}^{+}|\hbar v_\bk^i|\psi_{\bk}^{-}\rangle$ is the 
interband momentum matrix element with the velocity operator along 
$i$-direction defined as $ \hbar v_\bk^i = \partial_{k_i} H_\bk$, 
see Eq. (\ref{current}). For the mean-field Hamiltonian $H_\bk$, Eq. 
(\ref{MeanField}), we finally obtain
\begin{flalign}
P_\bk^i=  -t \left|\phi_\bk^{ee}\right|\left[ i\partial_{k_i}\varphi_\bk+
\sin\vartheta_\bk\partial_{k_i}m_\bk\right]\;.
\label{px}
\end{flalign} 

If $m_\bk=0$,  
\begin{flalign}
|P_\bk^i|^2 =  \frac{|\phi_\bk^{ee}|^2}{|\phi_\bk|^2}|P_\bk^{i,0}|^2   
\label{psq}
\end{flalign}
\noindent
where $|P_\bk^{i,0}|^2=\frac{t^2 a^2}{16}g_\bk$ corresponds to the value in 
non-interacting case and 
\begin{flalign}
g_\bk = 18 + 4|\phi_\bk|^2 -24 \Re{\tilde{\phi}_\bk} + 
18\frac{[\Re{\tilde{\phi}_\bk}]^2 - [\Im{\tilde{\phi}_\bk}]^2}{|\phi_\bk|^2}. 
\end{flalign}
with $\tilde{\phi}_\bk =e^{-i\bm {k}\cdot\bm{\delta}_3}\phi_\bk$.\cite{Peres08}
 
\subsection{Optical conductivity}
There has been considerable work on the renormalization of the conductivity due to electron-electron interaction. To leading order in $\alpha^*=\alpha\frac{v_F}{v_F^*}$, one gets via a renormalization group analysis
\begin{align}
\frac{\sigma}{\sigma_0}=1+C\alpha^*+O({\alpha^*}^2)
\label{SigmaDiracSI}
\end{align}
with $C_1=25/12-\pi/2\approx0.51$ (hard cut-off),\cite{Herbut08} $C_2=19/12-\pi/2\approx0.01$ (soft cut-off),\cite{Mishchenko08,Sheehy09} and $C_3=11/6-\pi/2\approx0.26$ (dimensional regularization).\cite{Juricic10} Chiral anomalies have been claimed to be responsible for these discrepancies and a perturbative analysis based on the tight-binding model yields $C=C_3$.\cite{Rosenstein13} 

Since for particle-hole symmetric {Hamiltonian} there is no sign-problem, the optical conductivity of graphene can also be studied using quantum Monte Carlo calculations and values $C\approx0.05$ were obtained for different interaction strengths.\cite{Boyda16} Large error bares were attached to these calculations, but we have implemented an improved scheme for the analytical continuation, obtaining significantly more precise results (see below).

From our Hartree-Fock calculations, we obtain the analytical result $C=0.25$ for unscreened interaction,  but numerical results suggest lower values for self-screened interaction with $C\approx0.05$ for suspended graphene ($\alpha=2.5$ for $t=2.7$eV).

For $m_\k=0$ and momenta close to the Dirac point, we have $|\phi_\k|=\frac{3}{2}ak$ and an isotropic dispersion. The regular part of the real optical conductivity then reads:
\begin{align}
\sigma(\omega=\mathcal{E}_\k^+)=\sigma_0\frac{\mathcal{E}_\k^+}{k\partial_k\mathcal{E}_{\k}^+}
\end{align}
As shown by Sharma and Kopietz,\cite{Sharma16} close to the neutrality point we have the following functional behavior valid to all orders in $\alpha$:
\begin{align}
v_F^*=v_F(A+B\ln(\Lambda/k))
\end{align}
From  Eq. (\ref{condition1}), we get $A=1$ and integrating the above scaling law yields $\mathcal{E}_\k^+=\hbar v_Fk(1+B\ln(\Lambda/k)+B)$. For small $\omega$, we then get  the following result:
\begin{flalign}
\sigma=\sigma_0\left(1+\frac{Bv_F}{v_F^*}\right)\;,   
\label{SigmaDirac2}
\end{flalign}
with $\sigma_0=\frac{g_sg_ve^2}{16\hbar}$ the universal conductivity. For unscreened interaction, we have the analytical result $B=\alpha/4$ and thus $C=0.25$. For screened interaction, a good approximation is given by $B=\alpha/4\epsilon^{RPA}$ and thus $C(\alpha)=0.25/(1+\frac{\pi}{2}\alpha)$. These results also holds in the Dirac cone approximation.  

\subsection{Drude weight} 
The optical $f$-sum rule reads   
\begin{flalign}
\int_0^\infty \Re{\sigma}_{ii}(\omega) d\omega = -\frac{\pi}{2}\frac{\langle j_D^i\rangle}{A}
\end{flalign} 
\noindent
where $j_D^i\equiv-\frac{e^2}{\hbar^2}\sum_{\bk}\partial_{k_i}^2H_\bk$ is the diamagnetic current operator.  From 
the above $f$-sum rule we obtain the expression for the Drude weight: 
\begin{flalign}
D_{ii}= 
-\frac{2}{\pi}\int_0^\infty {\sigma_{ii}^{reg}}(\omega) d\omega-\frac{\langle j_D^i\rangle}{A}
\end{flalign} 

Because of the knowledge of the HF-wave function, we can derive an analytical formula for the Drude weight in the presence of electron-electron interaction. With the spin-degeneracy $g_s=2$, we obtain the general result, valid also in the presence of an arbitrary mass term,
\begin{flalign}
D_{ii} = \left(\frac{e }{\hbar}\right)^2\frac{g_s}{A}\sum_{\bk,s} 
[\partial_{k_i}^2\mathcal{E}_{\bk}^s] f(\mathcal{E}_{\bk}^s)\;.
\label{drudeSI}
\end{flalign}
Note that the expectation value of the diamagnetic 
current operator, $\langle j_D^i\rangle$, is not equal to the Drude weight since the 
Hellmann-Feynman theorem does not apply for the second derivative. For 
particles with mass $m$ and parabolic dispersion, the well-known Drude weight 
$D=e^2n/m$ with $n$ the particle density is obtained.\cite{Stauber14} The above 
formula generalizes this result to two bands and we expect it to hold also for 
general multi-band systems (index $s=1...n$).

Using partial integration, one gets at $T=0$ in the thermodynamic limit 
$A_s\to\infty$
\begin{flalign}
D_{ii}&=g_s\left(\frac{e}{h}\right)^2\int 
dk_{\bar{i}}\sum_{k_x|E_{\bk}^s=\mu,s}|\partial_{k_i}E_{\bk}^s|\rightarrow\left(\frac{e}{h}\right)^2\pi\mu\;,
\end{flalign}
where the last equation holds for isotropic Fermi surfaces.We thus obtain a direct link between the Drude weight and the Fermi energy $\mu$. Comparing systems with the same density, i.e., with the same Fermi wave number $k_F$,  we obtain the same renormalization of the Drude weight as for the Fermi velocity:
\begin{flalign}
\frac{D}{D_0}=\frac{v_F^*}{v_F}\;.
\end{flalign}
The effect of temperature, large doping and disorder might lead to important changes which can be discussed in the presented formalism.
\section{Numerical implementation and results}
The Hartree-Fock equations with unscreened interaction are numerically solved using a grid with up to  $N_c=15000^2$ lattice sites. This allows us to accurately discuss the scaling behavior for the undoped system as well as the Fermi-velocity renormalization for small finite electronic densities down to $n=10^{10}$cm$^{-2}$. 

In order to match the experimental results, it is crucial to incorporate self-screening effects. For this, the static polarization function {$\chi_{\G,\G'}(\q)$ has} been calculated for momenta on a small grid $N_c=300^2$ and later we used bilinear interpolation to obtain the polarization on a larger grid $N_c=3000^2$. The momentum summation for the polarization function has also been performed on a larger grid $N_c=600^2$. Finally, we approximated $\chi$ for small momenta by the analytical formula of the Dirac cone approximation, see Eq. (\ref{chiLocalField}). In both cases, i.e., for the unscreened as well as for the screened interaction, the sum and matrix over the reciprocal lattice vectors $\G=n\bm {b}_1+m\bm {b}_2$ is truncated by $|n|,|m|\leq n_{max}$ with $n_{max}=4,6$.
   
For the self-consistent solution, we approximated the static polarization by the bare polarization divided by the renormalized Fermi velocity. This is justified following the work by Sodemann and Fogler\cite{Sodemann12} who found that $\epsilon(q)=1+\frac{\pi}{2}\alpha^*+O({\alpha^*}^2)$ where $\alpha^*=\frac{1}{4\pi\epsilon_0\epsilon}\frac{e^2}{\hbar v_F^*}$ is the fine-structure constant with respect to the renormalized velocity. The self-consistent dispersion is scaled by a factor 1.4 compared to the dispersion coming from the bare polarization and is usually larger than the experimental data. 

\subsection{Electronic dispersion}
In Fig. \ref{UnScreened}, we show the universal scaling behavior {$(\frac{v_F^*}{v_F}-1)/\alpha$} for different grid sizes $N_c=3000^2,15000^2$. A fit to
\begin{align}
v_F^*=v_F(1+B\ln(\Lambda/k))
\label{FitVF}
\end{align}
with $B=\alpha/4$ yields $\Lambda=1.75\,\mathrm{\mathring{A}}^{-1}$. Screening can be incorporated by $\alpha\rightarrow\alpha/\epsilon^{RPA}$ which agrees well with the full calculation. 

For finite densities, we observe a kink in $v_F^*$ at $k_F$ which is smeared out at finite temperatures. This kink is clearly an artifact of the unscreened interaction and screening cannot be incorporated afterwards. It is thus crucial to calculate the Fermi velocity renormalization including the self-screened interaction. The results are shown in Fig. \ref{Screened} for the two orthogonal directions of the Bloch momentum. As expected, finite doping acts as a cutoff of the scaling law, but sizable effects are still observed if we trace the Fermi velocity at the Fermi level $k_F$. We also obtain a larger prefactor of the logarithm (black dashed line) compared to the undoped case (magenta dotted-dashed line) along the $KM$-direction.

Finally, in Fig. \ref{ScreenedFit}, we present the numerical curves for the renormalized Fermi velocity at the neutrality point for various coupling strengths $\alpha$. We used $t=3.1$eV, but the asymptotic results do not depend on the hopping parameter. The fits to Eq. (\ref{FitVF}) are shown as blue lines. They are slightly different for the two orthogonal directions, and we will consider the one along the $K\Gamma$-direction since it is more close related to Dirac cone physics.\cite{Stauber10a,Stauber10b}

\begin{figure}
\includegraphics[width=0.49\columnwidth]{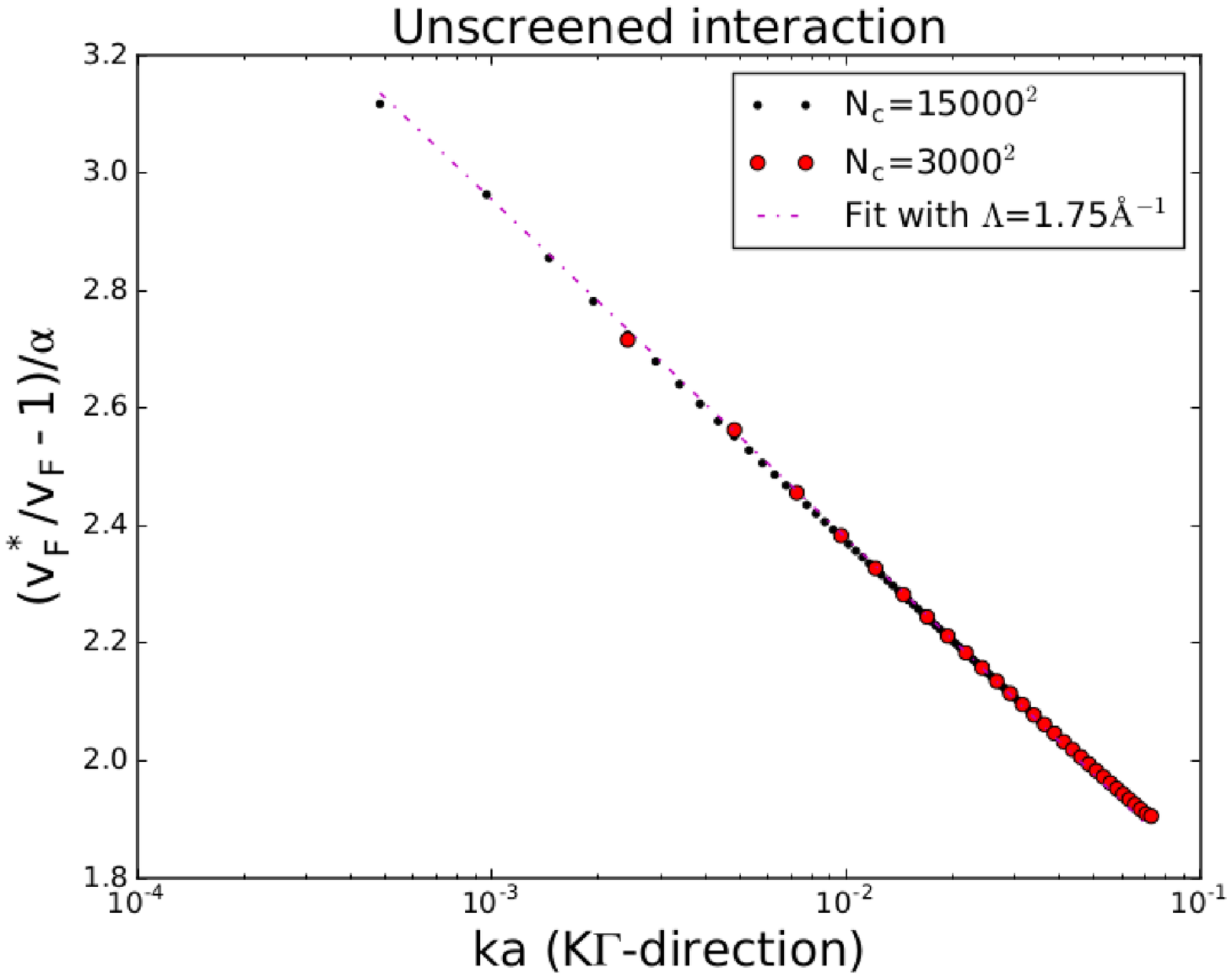}
\includegraphics[width=0.49\columnwidth]{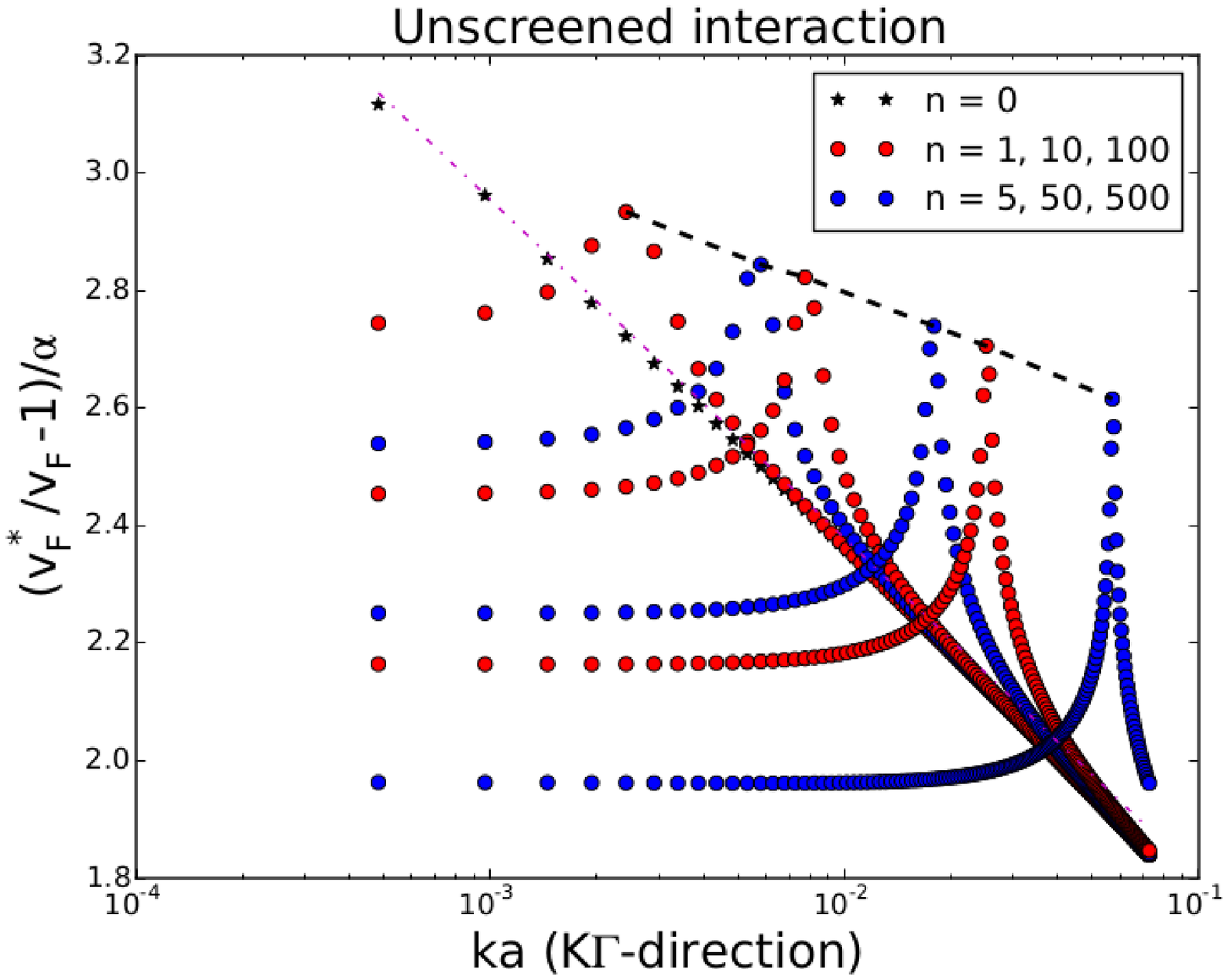}
\caption{(color online): Left: Universal expression of the renormalized Fermi velocity screening as a function of the momentum for two different grids $N_c=3000^2,15000^2$. Also shown the fit to Eq. (\ref{FitVF}) with $B=\alpha/4$. Right. Universal expression of the renormalized Fermi velocity screening as a function of the momentum for different electronic densities $n$ in units of $10^{10}$cm$^{-2}$. Also shown the fit for $n=0$ (magenta dashed-dotted line) and the Fermi velocity at $k_F$ (black dashed line).}
\label{UnScreened}
\end{figure}

\begin{figure}
\includegraphics[width=0.49\columnwidth]{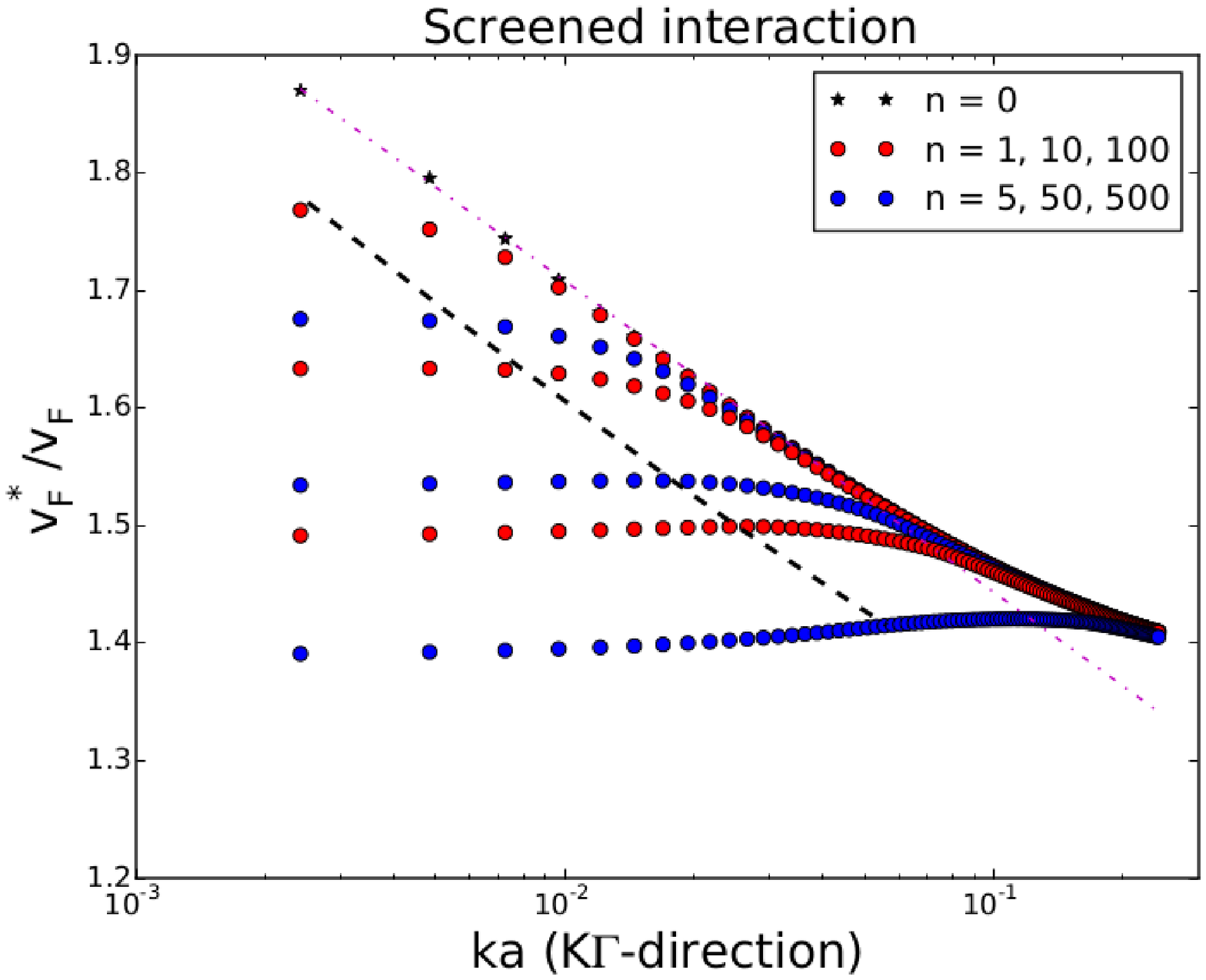}
\includegraphics[width=0.49\columnwidth]{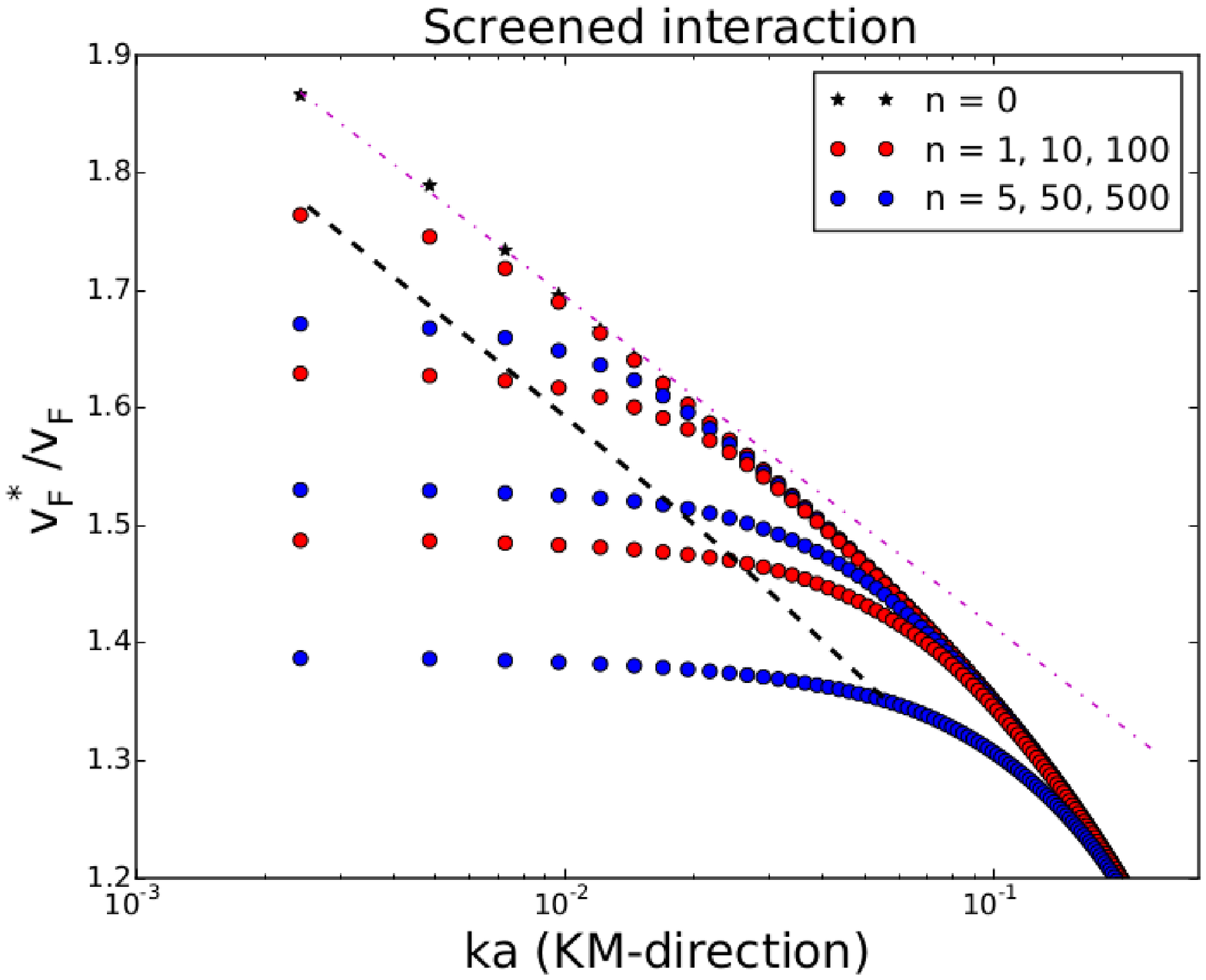}
\caption{(color online): Renormalized Fermi velocity for suspended graphene ($\alpha=2.2$ with $t=3.1$eV) screening as a function of the momentum for the two orthogonal directions and for different electronic densities $n$ in units of $10^{10}$cm$^{-2}$. Also shown the fit for $n=0$ (magenta dashed-dotted line) and the Fermi velocity at $k_F$ (black dashed line).
}
\label{Screened}
\end{figure}

\begin{figure}
\includegraphics[width=0.49\columnwidth]{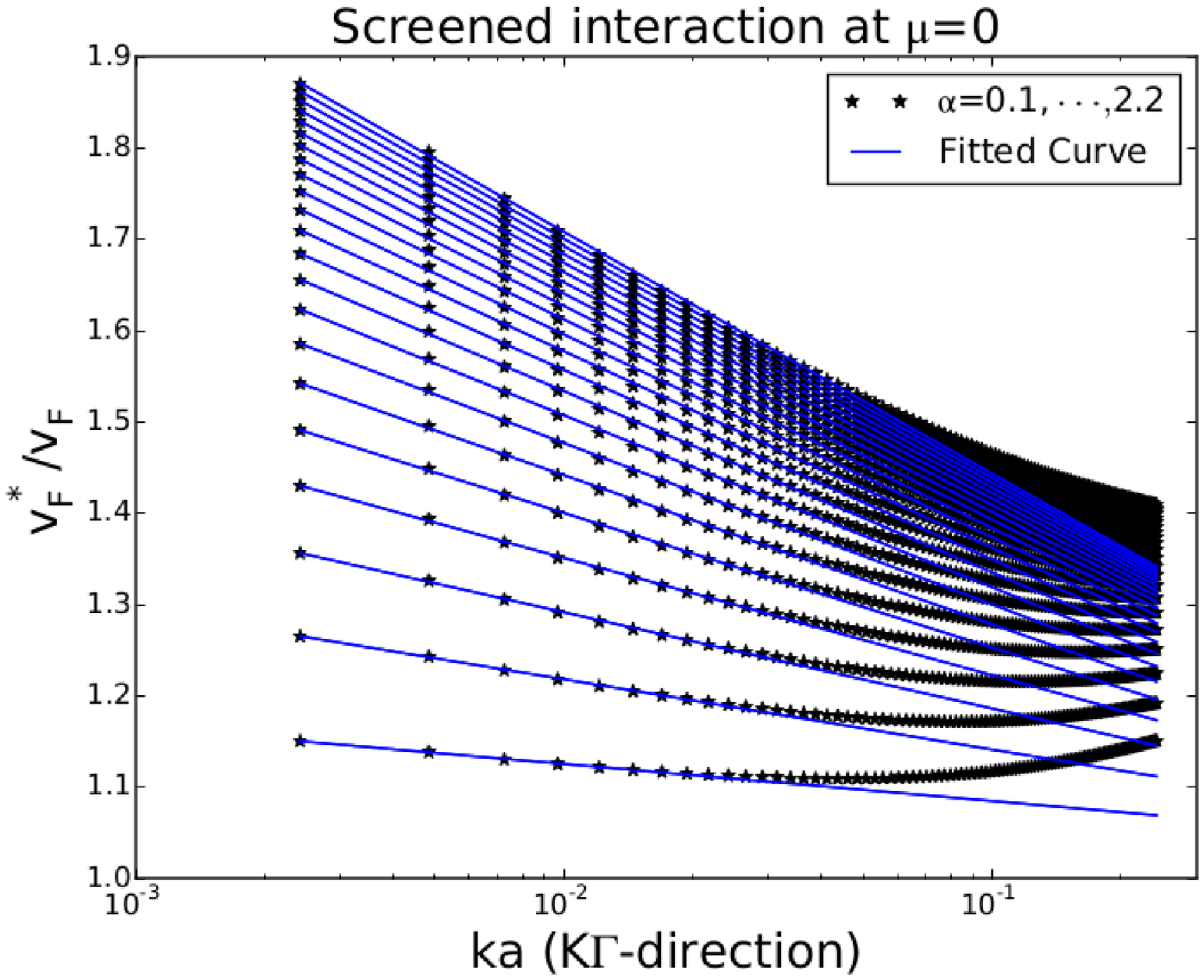}
\includegraphics[width=0.49\columnwidth]{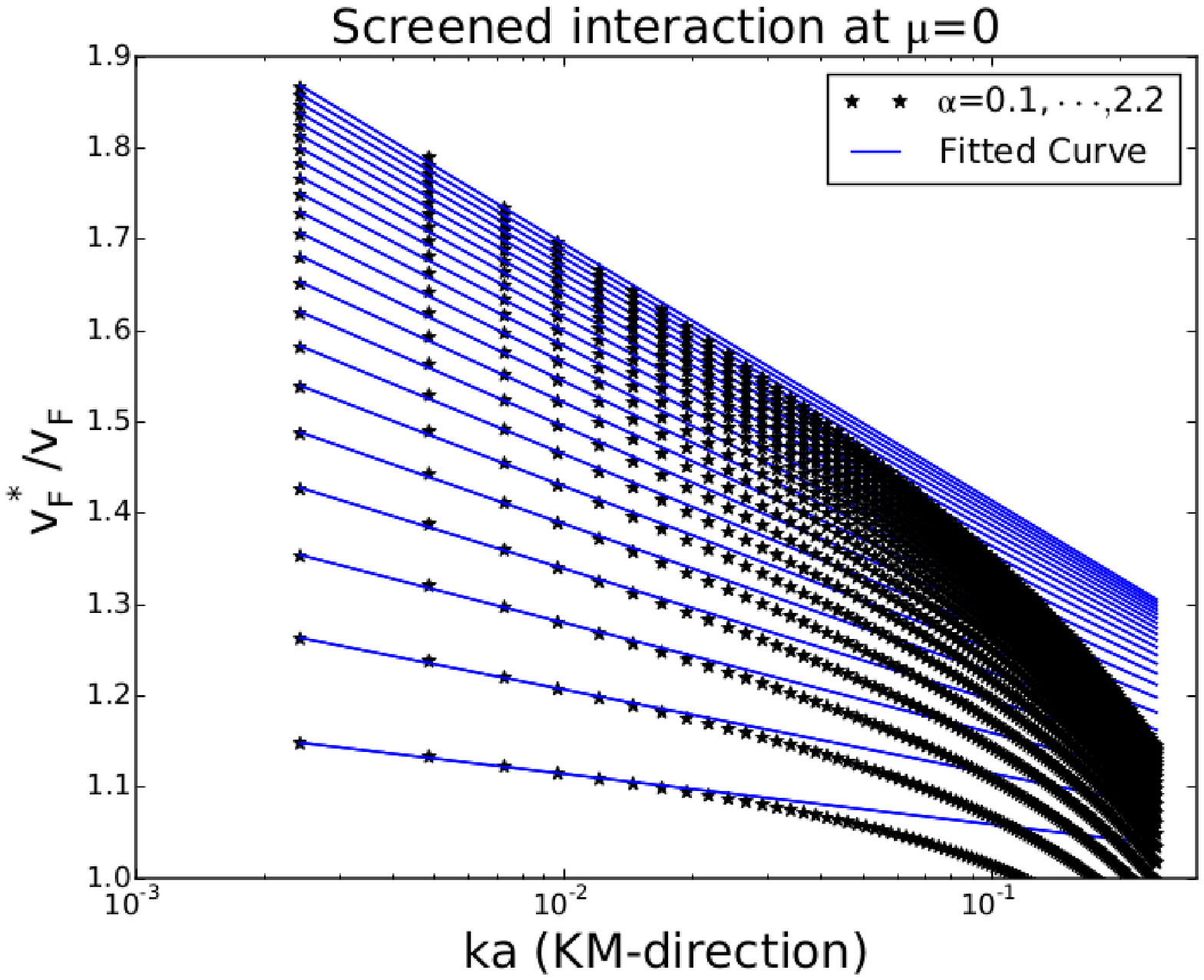}
\caption{(color online): Renormalized Fermi velocity at the neutrality point ($\mu=0$) as a function of the momentum for the two orthogonal directions and for various coupling constants $\alpha=0.1,\dots2.2$ with $t=3.1$eV. Also shown the fit functions to Eq. (\ref{FitVF}).
}
\label{ScreenedFit}
\end{figure}

\subsection{Optical conductivity}
\begin{figure}
\includegraphics[width=0.9\columnwidth]{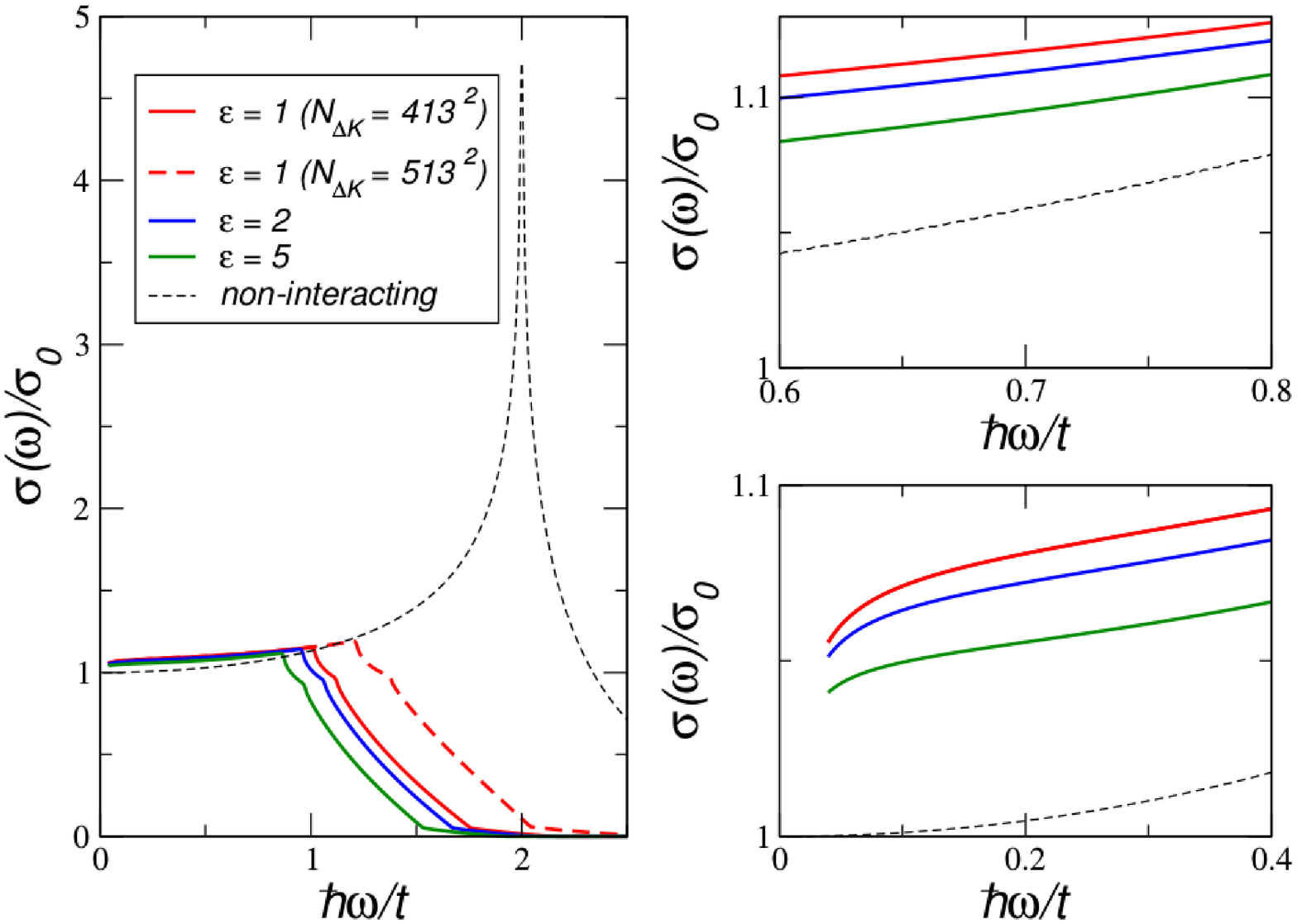}
\caption{(color online): The optical conductivity for suspended graphene with $t=2.7$eV calculated around one Dirac point including $N_{\Delta K}=413^2$ (solid red curve) and $N_{\Delta K}=513^2$ (dashed red curve) grid points based on a system with $N_c=3000^2$ lattice sites. This is compared to the optical conductivity of graphene on a substrate with $\epsilon=2$ or $\alpha=1.25$ (blue) and $\epsilon=5$ or $\alpha=0.5$ (green), calculated for $N_{\Delta K}=413^2$. Also shown the optical conductivity without Coulomb interactions (dashed line)
}
\label{Sigma}
\end{figure}

In Fig. \ref{Sigma}, we show the optical conductivity for suspended graphene with $t=2.7$eV in the vicinity of the Dirac point, based on a system with $N_c=3000^2$ lattice sites. Convergence is reached for energies up to $\hbar\omega\approx1.2t$ for $N_{\Delta K}=513^2$ grid points of the Brillouin zone around one Dirac cone (due to time-reversal symmetry it is sufficient to only consider one Dirac cone). We also show the optical conductivity for graphene on a substrate with $\epsilon=2$ or $\alpha=1.25$ (blue) and $\epsilon=5$ or $\alpha=0.5$ (green) for $N_{\Delta K}=413^2$. 

In order to compare the mean-field results with the quantum Monte Carlo calculations, we choose an energy of $\hbar\omega=0.7t$ where the plateau of the conductivity has developed (see below). We obtain for the interacting conductivity $\sigma^* =1.117$ for $\epsilon=1$, $\sigma^*=1.109$ for $\epsilon=2$, and $\sigma^*=1.095$ for $\epsilon=5$. 
These values show excellent agreement with the improved quantum Monte Carlo calculations, see below.

\section{Precise Quantum Monte Carlo study of graphene conductivity}

We perform Quantum Monte Carlo (QMC) calculations for the following interacting tight-binding model for electrons at $\pi$-orbitals:
\begin{eqnarray}
\label{tbHam1}
 \hat{H} = \sum_{\langle x,y \rangle,\sigma} -t \lr{ \hat{a}^{\dag}_{y, \sigma} \hat{a}_{x, \sigma} + h.c.}
 +
 \frac{1}{2} \sum_{x,y} V_{xy} \hat{q}_x \hat{q}_y.
\end{eqnarray}
Here $\sum_{\langle x,y \rangle}$ means the summation over all pairs of nearest-neighbor sites. $\hat{a}^{\dag}_{x, \sigma}$, $\hat{a}_{x, \sigma}$ are the creation/annihilation operators for electrons with spin-up and spin-down, respectively. $t$=2.7 eV is the hopping amplitude. $\hat{q}_x = \sum_{\sigma} \hat{a}^{\dag}_{x, \sigma} \hat{a}_{x, \sigma} - 1$ is the charge operator at site $x$. $V_{xy}$ is the electron-electron interaction potential. We use the potentials calculated with the constrained RPA method \cite{Wehling} for suspended graphene (see \cite{Ulybyshev13} for details). These potentials correspond to the case $\varepsilon=1.0$ in our calculations. For $\varepsilon \neq 1$ we uniformly rescale potentials at all distances by a factor of $1/\varepsilon$. Since we can treat only finite sample, we impose periodic spatial boundary conditions as in Refs.~\cite{Buividovich:2012nx,Ulybyshev13,Lorenz2014}.

The calculation of graphene conductivity is performed with the same numerical procedure as was already described in \cite{Boyda16}. We calculate Euclidean current-current correlator $G(\tau)$ using Hybrid Monte Carlo algorithm, which is especially advantageous for systems of large spatial extent in the presence of long-range interaction. Since this particular Hamiltonian \ref{tbHam1} doesn't lead to sign problem in QMC calculations, we obtain numerically exact results without any further physical assumptions. Optical conductivity $\sigma(\omega)$ is extracted from the Green-Kubo relation:
\beq
 G(\tau) = \int_0^\infty  \sigma(\omega) K(\tau, \omega) d\omega, \\
 K(\tau, \omega)=\frac {\omega \cosh (\omega(\beta/2 -\tau))} {\pi \sinh(\omega \beta/2)}.
\eeq
This integral equation is unstable thus we need a special numerical algorithm with some kind of regularization to find the stable solution. It's especially important in our case when the input data for correlator $G(\tau)$ is defined up to some statistical errors because it was obtained in statistical Monte Carlo procedure.  In general we rely on the Backus-Gilbert (BG) method proposed in \cite{Meyer}, which we already employed in \cite{Boyda16}.

In order to improve accuracy and reduce statistical and systematic errors, we switched to substantially lower temperatures in comparison with our previous study (0.0625 and 0.125 eV  in comparison with 0.5 eV in \cite{Boyda16}).  According to the general idea of Euclidean time formalism, it leads to increased number of time slices. Unfortunately, it's impossible to rely on the old version of the BG method in this case, since the regularization scheme used in \cite{Meyer} and \cite{Boyda16} doesn't work well for lattices with large ($\sim 100$) number of Euclidean time slices. Significant modifications of the algorithm were introduced in order  to increase it's stability and to improve the resolution in frequency. These modifications are discussed below alongside with the careful study of possible systematic errors. 

We start from the brief description of the BG method. The estimator of the spectral function $\sigma(\omega)$ is defined as the convolution of the exact spectral function $\tilde\sigma(\tilde\omega)$ with the resolution function $\delta(\omega,\tilde\omega)$.  
 \begin{equation} \label{ConvolutionDeltaFunction}
     \sigma(\omega) = \int^{\infty}_0 d{\tilde\omega} \delta(\omega,\tilde\omega) \tilde\sigma(\tilde\omega).
 \end{equation}
 The resolution function is defined as a linear combination of the kernel profiles:
 \begin{equation}
   \delta(\omega,\tilde\omega) = \sum_{j} q_j(\omega) K(\tau_j, \tilde\omega).
 \end{equation}
The coefficients $q_j(\omega)$ are determined by minimizing the width of resolution function around $\omega$: $\partial_{q_j} D = 0$, where  $D$ is defined as
 \begin{equation}
   D \equiv \int^{\infty}_0 d\tilde\omega (\tilde\omega-\omega)^2 \delta(\omega,\tilde\omega).
 \end{equation}
 In practice, we minimize the width imposing the normalization condition:
  \begin{equation}
 \int^{\infty}_0 d\tilde\omega \delta(\omega,\tilde\omega) = 1
  \end{equation}
 and sometimes the additional condition $\delta(\omega,0) = 0$. The second condition is introduced in order to eliminate the influence of Drude peak on the estimator for optical conductivity. The result of this minimization yields
 \begin{equation}
   q_j(\omega) =\frac{W^{-1}(\omega)_{j,k}R_k}{R_n W^{-1}(\omega)_{n,m}R_m},
 \end{equation}
 where 
 \begin{equation} \label{W}
   W(\omega)_{j,k} = \int^{\infty}_0 d\tilde\omega (\tilde\omega-\omega)^2 K(\tau_j,\tilde\omega) K(\tau_k,\tilde\omega),~R_n = \int^{\infty}_0 d\tilde\omega K(\tau_n,\tilde\omega). 
 \end{equation}
 The matrix $W$ is extremely ill-conditioned, with $C(W) \equiv \frac{\lambda_{\text{max}}}{\lambda_{\text{min}}} \approx O(10^{20})$. Thus, one needs to regularize the method in order to obtain stable results for given set of data $G(\tau_i)$. 
 
 Previous studies employing the BG algorithm \cite{Meyer,Boyda16} have used the regularization scheme based on the addition of the covariance matrix $C_{j,k}$ of the Euclidean correlator $G$ to the kernel (\ref{W}):
 \begin{equation}
  W(\omega)_{j,k} \to (1-\lambda) W(\omega)_{j,k} + \lambda C_{j,k}, 
 \end{equation}
 where $\lambda$ is a small regularization parameter. The method worked well in \cite{Boyda16}  for lattices with 20 Euclidean time steps, but we failed with this approach in the calculations at small temperatures where the typical number of time steps is of the order of 100. Moreover, there is an important type of calculations where we apply the analytical continuation to formally exact correlator computed for the free fermions. It should be done in order to check the validity of the method and to study the systematic errors.  Despite the fact that the numerical values of exact data contain only round-off errors, it's still impossible to use them in the Green-Kubo relation without regularization. Unfortunately, the covariance matrix is not  defined for this kind of data, thus we again need some alternative regularization methods.
 
 In this paper we used combination of the so-called Tikhonov regularization and averaging of the correlator over intervals in Euclidean time. The Tikhonov regularization is widely used for the ill-posed problems of the form $Ax=b$. In this method, one seeks a solution to the modified least-squares function
 \begin{equation}
   \text{min} \left( \Vert A x - b \Vert^2_2 + \Vert \Gamma x \Vert^2_2 \right),
 \end{equation}
 where $\Gamma$ is an appropriately chosen matrix. In the standard Tikhonov regularization we choose $\Gamma = \lambda {\bm 1}$, where $\lambda$ is again a small regularization parameter. As usually, for small $\lambda$ the solutions fit the data well but are oscillatory, while at large $\lambda$ the solutions are smooth but do not fit the data. 
 
 In practice the Tikhonov regularization is introduced during the inversion of the kernel matrix $W$, where we use the singular value decomposition (SVD):
 \begin{equation}
   W = U \Sigma V^{\top}, ~UU^{\top} = VV^{\top} = {\bm 1},
   \label{SVD1}
 \end{equation}
 where $\Sigma = \text{diag}(\sigma_1,\sigma_2,\dots,\sigma_N),~\sigma_1 \geq \sigma_2 \geq \cdots \geq \sigma_N$. The inverse is thus easily expressed as 
 \begin{equation}
  W^{-1} = V \Sigma^{-1} U^{\top},~\Sigma^{-1} = \text{diag}( \sigma^{-1}_1,\sigma^{-1}_2,\dots,\sigma^{-1}_N ). 
  \label{SVD2}
 \end{equation}
 Applying standard Tikhonov regularization simply modifies the matrix $\Sigma$ in the following way 
 \begin{equation} \label{SVDStandardTikhonov}
   \Sigma^{-1}_{i,j} \to \tilde{\Sigma}^{-1}_{i,j} = \delta_{ij}\frac{\sigma_i}{\sigma^2_i + \lambda^2}.
 \end{equation}
 Thus, one can see that the small eigenvalues which satisfy $\lambda \gg \sigma_i$ are smoothly cut off.
 
 Now let's turn to the second regularization technique, namely, the averaging of the correlator over some intervals in time. In this procedure, we take the correlator data $\{ G(\tau_i); ~i=0,1,\dots , N_{\tau}-1 \}$ and map this to a new set $\{ \tilde{G}(\tilde{\tau}_j); ~j=1,\dots , N_{\text{int}} \}$ where
\beq
  \bar{G}(\bar{\tau}_j) &\equiv& \frac{1}{\bar{N}_j} \sum^{\bar{N}_j}_{i=1} G(\tau^{(j)}_i),\\ N_{\tau} &=& \sum^{N_{\text{int}}}_{j=1} \bar{N}_j,~1 \leq \bar{N}_j < N_{\tau}.  
\eeq 
 Due to the linearity of the Green-Kubo relation, one can construct $\{ \bar{K}(\bar{\tau}_j); ~j=1,\dots , N_{\text{int}} \}$ in an analogous manner and use the new kernel and the new correlator in the BG method without any further modifications of the algorithm.  The averaging increases the signal-to noise ratio in the input data. From the point of view of the ill-defined inversion problem for the kernel matrix $W$, we reduce the matrix size thus reducing the number of extremely small numbers $\sigma_i$ in $\Sigma$ (see \ref{SVD1} and \ref{SVD2}) and making the SVD decomposition more stable.

  We use lattices with spatial sizes $24\times24$, $36\times36$, $48\times48$, $72\times72$ and $96\times96$ (the two largest sizes only in the test calculations for the free fermions). In the most of our QMC runs the temperature is equal to 0.125 eV with 80 steps in Euclidean time.  As it was shown in \cite{phase_diag} this discretization is good enough if we are far from antiferromagnetic phase transition.  In analytical continuation we use Tikhonov regularization with constant $\lambda=10^{-12}...10^{-11}$ and additional averaging over Euclidean time starting from the 10th step in time. The length of averaging intervals is equal to 10 time slices.
 
\begin{figure} 
        \centering
        \includegraphics[scale=0.5, angle=0]{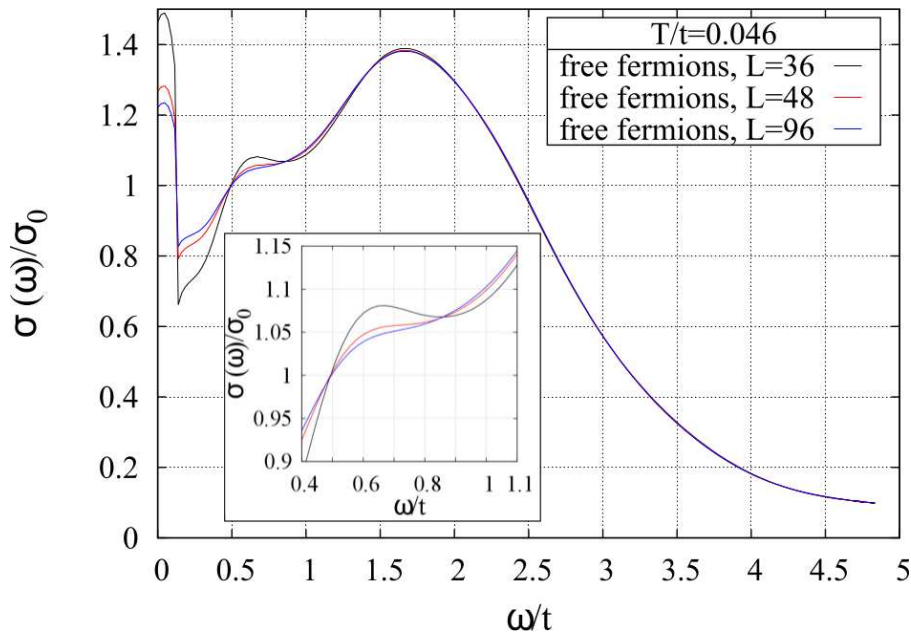}
        \caption{Finite size effects in the full profile  $\sigma(\omega)$ in the case of free fermions. Discontinuities at small frequency are the consequence of the presence of the Drude peak (delta-function for the free fermions) at $\omega=0$.}
        \label{fig:size_profiles}
\end{figure}
 
In the beginning we should study the systematic errors appeared due to finite lattice size, non-zero temperature and regularization during analytical continuation.
We start from the check that the
method applied to the free current-current correlator really
reproduces the analytic profile of conductivity for the free tight-binding model.
Fig. \ref{fig:size_profiles} shows the results of this study. First of all, one can see that the profiles become stable in the limit of large lattices. Special attention should be paid to the plateau with center around $\omega=0.7t$ (the point is defined as a position corresponding to the minimal value of the derivative $d\sigma / d\omega$).
In the limit of large  lattices this plateau reproduces correct value of 
 non-interacting conductivity at $\omega=0.7t$ ($\sigma_{NI}=1.058 
\sigma_0$) with systematic error less than one percent: see fig. \ref{fig:size_dep}.

\begin{figure} 
        \centering
        \includegraphics[scale=0.5, angle=-90]{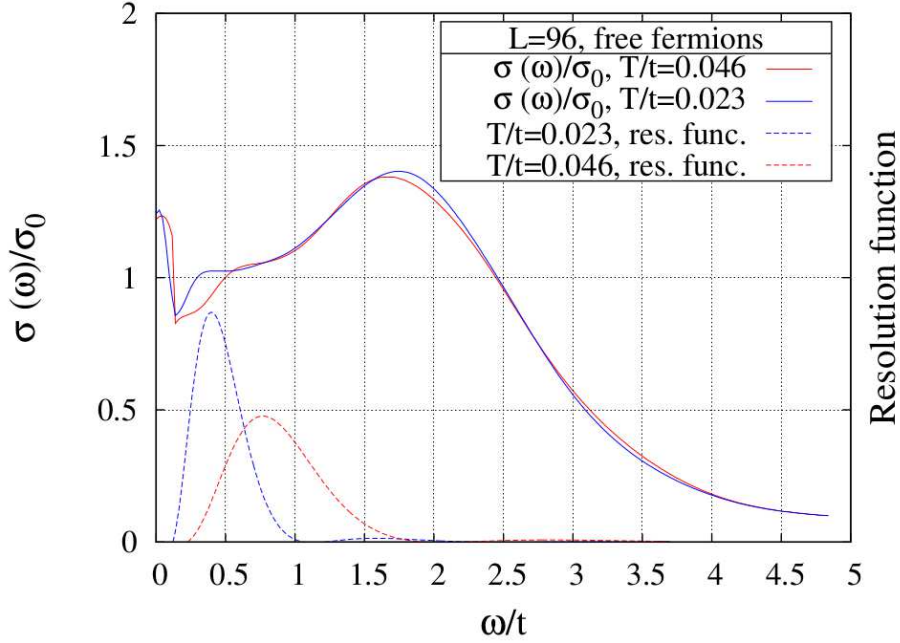}
        \caption{Comparison of $\sigma(\omega)$ profiles obtained from analytical continuation of free current-current correlator at two temperatures: T=0.0625 eV and T=0.125 eV. The resolution functions $\delta(\omega,\tilde \omega)$ with centers ($\omega$) at the plateau are plotted to illustrate the improvement of the resolution once we decrease the temperature. The plateau tends to correct value $\sigma=\sigma_0$ when temperature goes down.}
        \label{fig:comparison_temp}
\end{figure}

Another argument why we should look at this plateau is that it evolves into the plateau corresponding to the Dirac fermions in the low temperature limit. 
This feature is demonstrated in the figure \ref{fig:comparison_temp}. Indeed, once the temperature is decreased, the plateau shifts towards lower frequencies. The value of $\sigma (\omega)$ at plateau simultaneously goes towards the standard limit of 2D Dirac fermions $\sigma_0$. Thus we can conclude that the correct value of conductivity for Dirac fermions can be reproduced in our approach once we simultaneously increase the lattice size and decrease the temperature.

 \begin{figure} 
        \centering
        \includegraphics[scale=0.5, angle=-90]{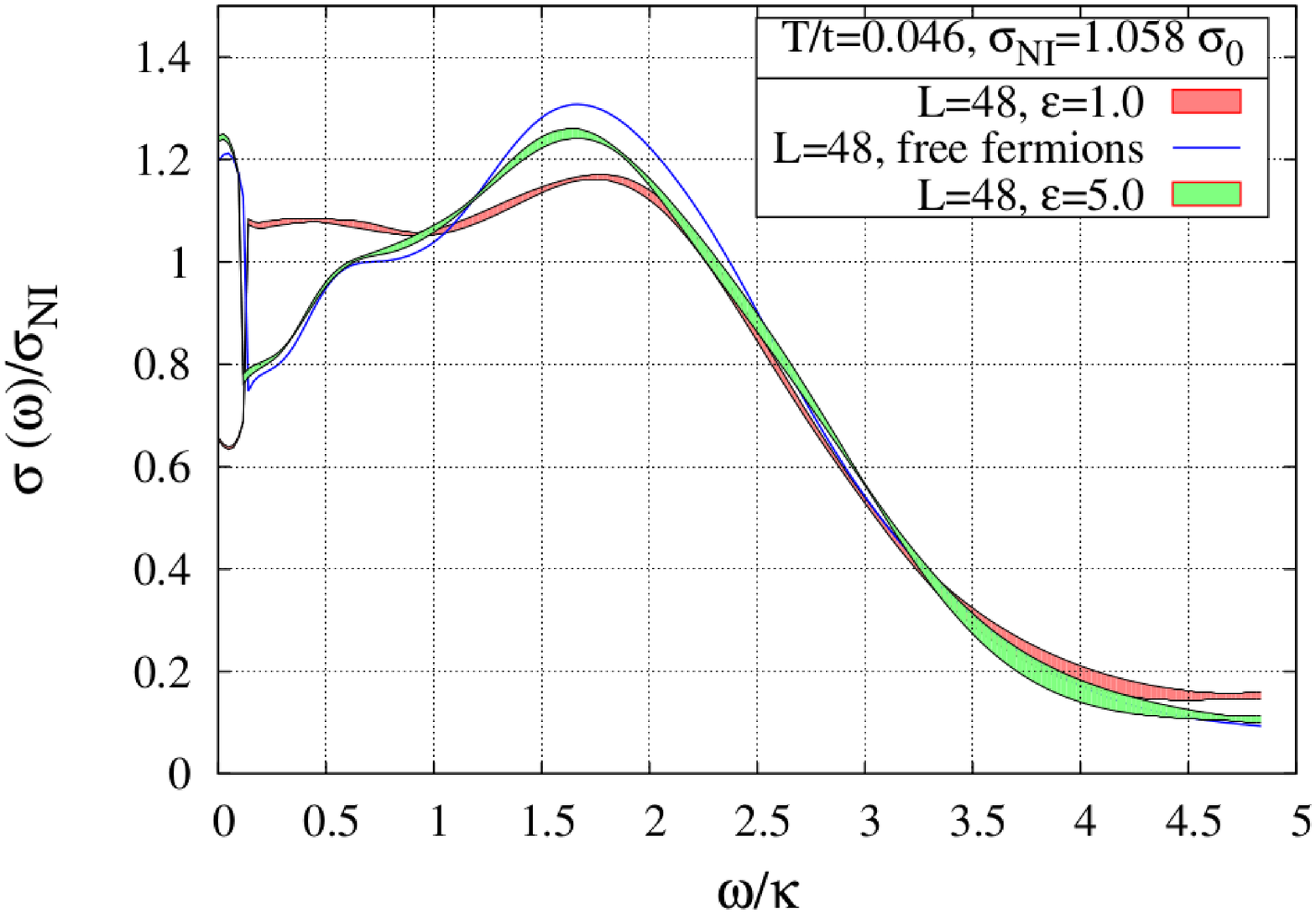}
        \caption{Comparison of interacting case (suspended graphene) with the results for the free fermions. Temperature is equal to 0.125 eV and lattice size  $L=48$ in both cases.}
        \label{fig:comparison48_80}
\end{figure}

However, in practice it's impossible to make QMC simulation at very large lattices and very small temperature (the latter needs increased number of steps in Euclidean  time). Thus we'll stop at lattice size equal to $48\times 48$, where correct value of $\sigma|_{\omega=0.7t}$ for the free fermions is already reproduced with good precision once we perform calculations at the temperature T=0.125 eV. 
This frequency is used in all further calculations in the paper where we compare QMC results with analytic calculations.

The last potential source of systematic errors is the regularization. We checked the dependence of conductivity on the constant $\lambda$ in the whole interval $\lambda=10^{-12}...10^{-11}$ and the relative variation of conductivity $\sigma|_{\omega=0.7t}$ is smaller than 1\%.  
Thus we can conclude that systematic errors are under control since all of them are actually smaller than statistical uncertainties. 

Conductivity profiles obtained in real QMC calculations are shown in the figure \ref{fig:comparison48_80}. Once can see that the general feature of $\sigma(\omega)$ profiles: the plateau around $\omega=0.7 t$, is preserved. Actually, for suspended graphene the plateau is even substantially wider.  Results for conductivity at $\omega=0.7t$ are summarized in the figure \ref{fig:size_dep}. Interestingly, the finite size effects for interacting systems become milder with increased interaction strength. This feature provides additional support for the statement that the finite size effects are under control, because the case of free fermions shows the upper limit for the finite size effects.
Results for the largest accessible in QMC lattice ($48\times48$) are used in the main text for comparison with other approaches.

 \begin{figure} 
        \centering
        \includegraphics[scale=0.5, angle=-90]{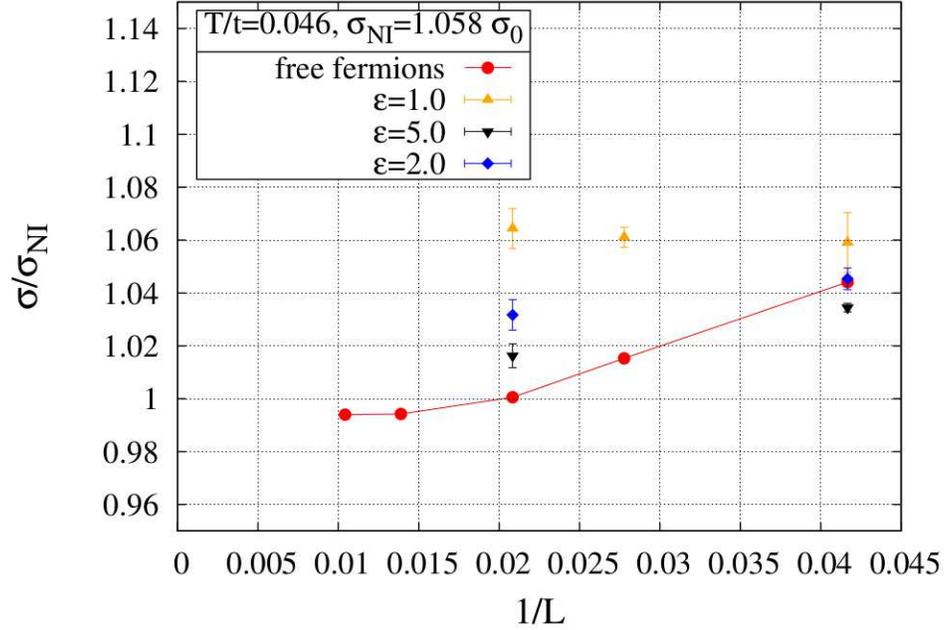}
        \caption{Dependence of optical conductivity $\sigma|_{\omega=0.7t}$ on the lattice size in the free case and in interacting systems. T=0.125 eV in all calculations.}
        \label{fig:size_dep}
\end{figure}
\end{widetext} 
\end{document}